\pdfoutput=1

\documentclass[final,3p,times]{elsarticle}

\usepackage{tikz}
\usepackage{bm}
\usepackage{amsmath}
\usepackage{mathtools}
\usepackage{dirtytalk}
\usepackage{derivative}
\usepackage{threeparttable}
\usepackage{lscape}
\usepackage{float}
\usepackage[pagewise]{lineno}
\usepackage{booktabs}
\usepackage{threeparttable}
\usepackage{caption}
\usepackage{pgfplots}
\usepackage{subcaption}
\usepackage[short,nocomma]{optidef}
\usepackage{hyperref}
\usepackage{graphicx}
\usepackage{afterpage}
\usepackage{array}
\usepackage{placeins}
\usepackage{dblfloatfix}
\usepackage{multirow}
\usetikzlibrary{shapes.geometric, arrows}

\pgfplotsset{compat=1.18}
\newcommand{\tran}{{\mkern-1.5mu\mathsf{T}}}
\biboptions{sort&compress}

\begin{document}

\begin{frontmatter}

\author[0]{Sebastián Espinel-Ríos\corref{cor1}}
\ead{sebastian.espinelrios@csiro.au}

\author[1,2,3,4]{José L. Avalos}

\author[5]{Ehecatl Antonio del Rio Chanona}

\author[6]{Dongda Zhang}

\cortext[cor1]{Corresponding author}

\affiliation[0]{organization={Biomedical Manufacturing Program, Commonwealth Scientific and Industrial Research Organisation},
            city={Clayton},
            country={Australia}}

\affiliation[1]{organization={Department of Chemical and Biological Engineering, Princeton University},
            city={Princeton},
            country={United States}}

\affiliation[2]{organization={Omenn-Darling Bioengineering Institute, Princeton University},
            city={Princeton},
            country={United States}}

\affiliation[3]{organization={The Andlinger Center for Energy and the Environment, Princeton University},
            city={Princeton},
            country={United States}}

\affiliation[4]{organization={High Meadows Environmental Institute, Princeton University},
            city={Princeton},
            country={United States}}

\affiliation[5]{organization={Department of Chemical Engineering, Imperial College London},
            city={London},
            country={United Kingdom}}
            
\affiliation[6]{organization={Department of Chemical Engineering, University of Manchester},
            city={Manchester},
            country={United Kingdom}}

\title{Reinforcement learning for efficient and robust multi-setpoint and multi-trajectory tracking in bioprocesses}

\begin{abstract}
{Efficient and robust bioprocess control is essential for maximizing performance and adaptability in advanced biotechnological systems. In this work, we present a reinforcement-learning framework for multi-setpoint and multi-trajectory tracking. Tracking multiple setpoints and time-varying trajectories in reinforcement learning is challenging due to the complexity of balancing multiple objectives, a difficulty further exacerbated by system uncertainties such as uncertain initial conditions and stochastic dynamics. This challenge is relevant, e.g., in bioprocesses involving microbial consortia, where precise control over population compositions is required. We introduce a novel return function based on multiplicative reciprocal saturation functions, which explicitly couples reward gains to the simultaneous satisfaction of multiple references. Through a case study involving light-mediated cybergenetic growth control in microbial consortia, we demonstrate via computational experiments that our approach achieves faster convergence, improved stability, and superior control compliance compared to conventional quadratic-cost-based return functions. Moreover, our method enables tuning of the saturation function’s parameters, shaping the learning process and policy updates. By incorporating system uncertainties, our framework also demonstrates robustness, a key requirement in industrial bioprocessing. Overall, this work advances reinforcement-learning-based control strategies in bioprocess engineering, with implications in the broader field of process and systems engineering.}
\end{abstract}

%%Graphical abstract
\begin{graphicalabstract}
\centering
\includegraphics[scale=0.5]{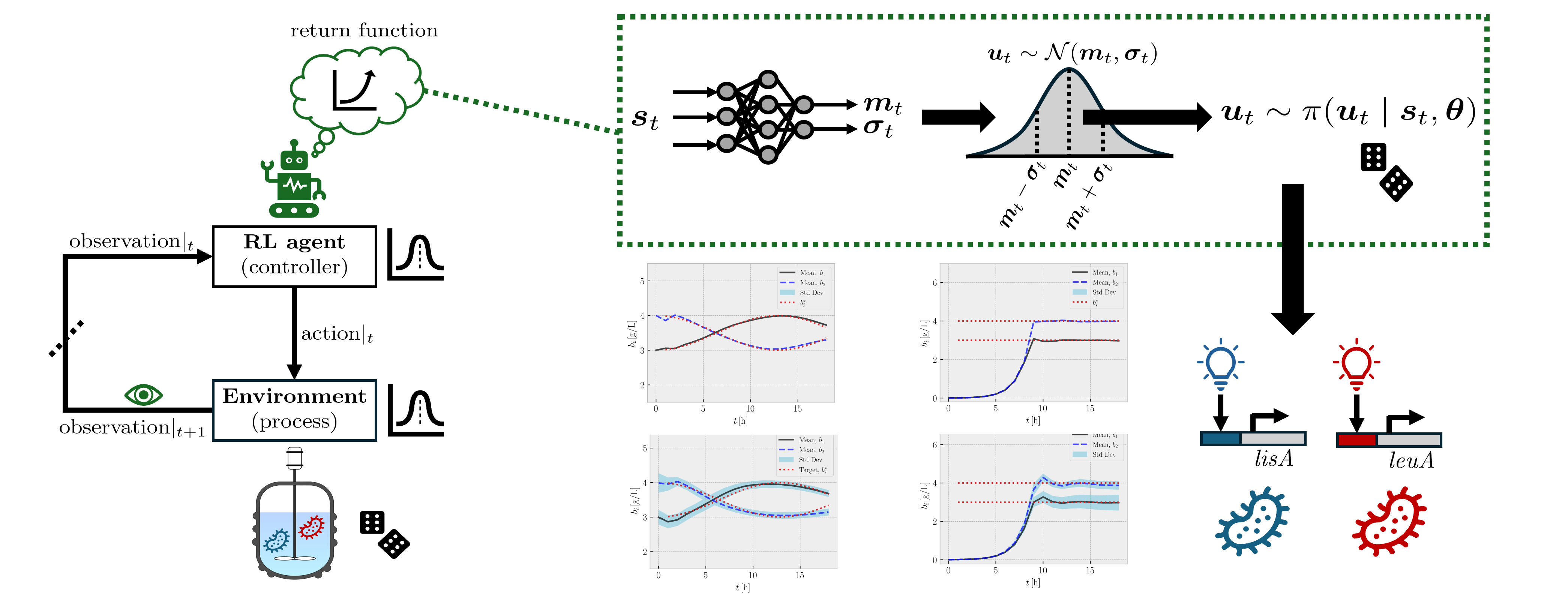}
\end{graphicalabstract}

%%Research highlights
\begin{highlights}
\item Reinforcement learning tailored for multi-setpoint and multi-trajectory tracking.
\item A novel return function enhances learning stability, convergence, and control performance.
\item Proposed return function based on multiplicative reciprocal saturation functions.
\item Framework accounts for system uncertainties, ensuring robust bioprocess control.
\item Computational experiments involving cybergenetic growth control in microbial consortia.
\end{highlights}


\end{frontmatter}

\section{Introduction}
\label{sec:introduction}
Bioprocesses involve the use of microorganisms to catalyze the production of value-added products through cellular metabolic networks, thereby contributing to sustainability and the bioeconomy \cite{nielsen_innovation_2022}. Metabolic engineering, which typically relies on genetic engineering interventions, plays a crucial role in maximizing production efficiency in biotechnology \cite{ko_tools_2020, hartline_dynamic_2021}. {However, maintaining redox balance, net ATP production, and thermodynamic feasibility simultaneously in engineered metabolic pathways, while also minimizing resource burden and properly managing intrinsic metabolic trade-offs, is often a very challenging task \cite{tian_refactoring_2020,MAO2024108401}}.

Biotechnological processes involving microbial consortia have received increasing attention in recent years due to the numerous possibilities they offer for bioproduction (cf. e.g., \cite{JIANG20231430, darvishi_applications_2024}). For instance, complex metabolic pathways can be split among different consortium members, reducing the metabolic burden on individual cells, an approach known as division of labor. Additionally, the inherent biological properties of specific engineered cells or species can be harnessed for targeted transformations, such as better expression of certain plant enzymes by yeasts. A major challenge, however, lies in the efficient operation and optimization of consortia, as the fastest-growing member in the bioreactor will eventually dominate in the absence of appropriate controllers or engineered co-dependencies.

Traditionally, bioprocesses have been optimized and operated largely through empirical or heuristic approaches, often relying on so-called \textit{golden-batch} recipes. While some feedback control strategies, such as Proportional-integral-derivative (PID) control \cite{astrom_feedback_2021}, are commonly used in commercial bioreactors for setpoint tracking of environmental variables like pH, temperature, and dissolved oxygen \cite{JONES2023209}, these regulate only \textit{lower-level} operational parameters. PID control is considered \textit{reactive}, as it applies proportional, integral, and derivative gains based on error measurements without anticipating or predicting the plant’s future behavior. Moreover, PID control is designed for linear systems, limiting its flexibility in handling more complex nonlinear dynamics.

There have been significant advances in feedback control strategies for bioprocesses that regulate \textit{higher-level} process dynamics, involving biomass, substrate, and product concentration profiles (cf. e.g., \cite{zupke_mpc,CRAVEN2014344,espinel_met_cyberg,espinel-rios_hybrid_2024}). For instance, model predictive control (MPC) updates control actions by solving open-loop optimal control problems constrained by a (nonlinear) dynamic system model, the plant’s measured or estimated states, and possibly additional (nonlinear) system constraints \cite{rawlings_model_2020}. Although MPC can handle \textit{sufficiently small disturbances}, it relies on a predefined model that does not inherently adapt over time. Some variations incorporate model adaptation \cite{ADETOLA2009320,JABARIVELISDEH2020106744}, but determining which model components to recalibrate is not trivial. Additionally, \textit{nominal} MPC is deterministic and does not explicitly account for stochastic behavior, which requires more advanced formulations, such as stochastic or robust MPC \cite{rawlings_model_2020}.

\begin{figure*} [htb!]
\begin{center}
\includegraphics[scale=0.35]{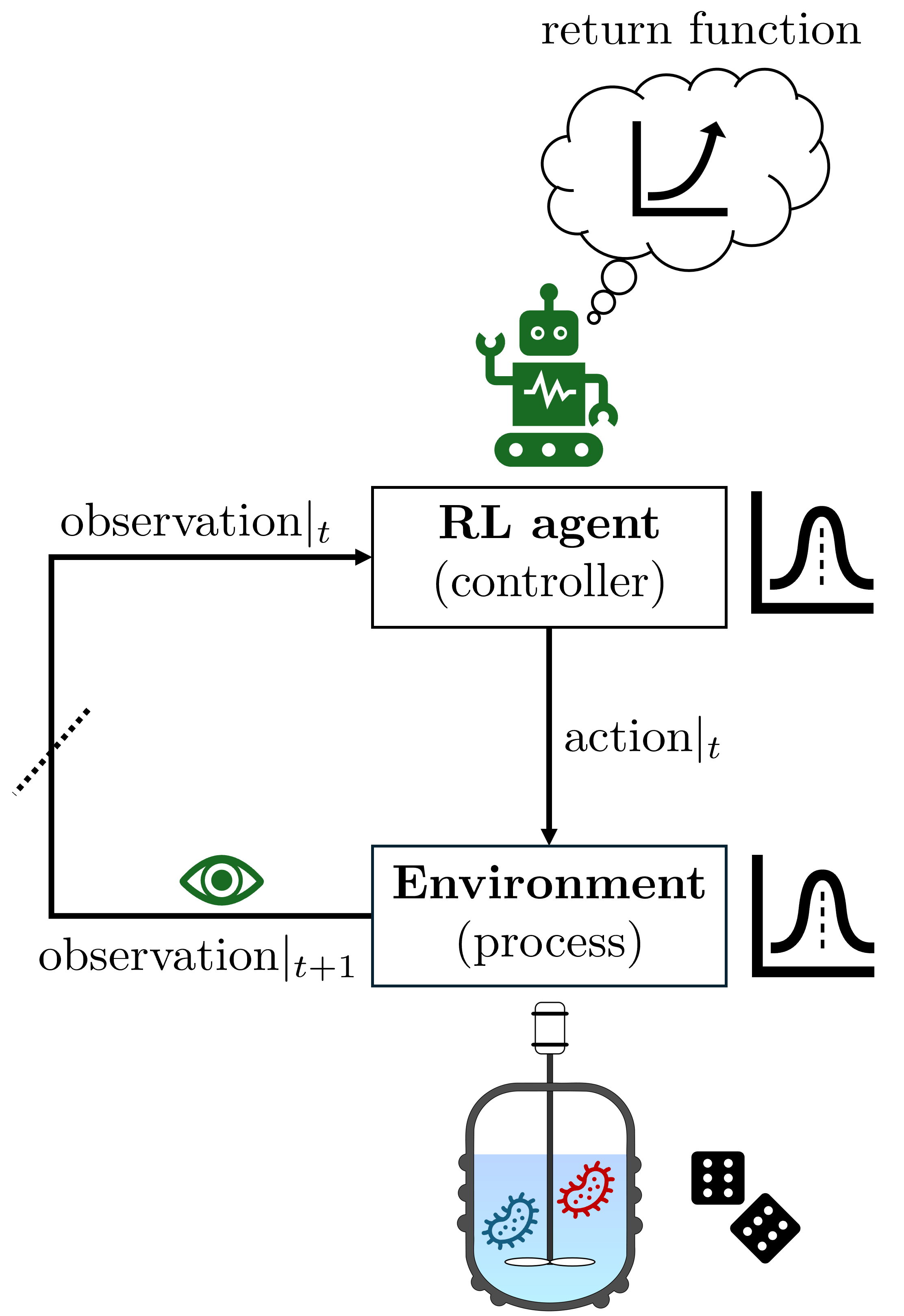}
\end{center}
\caption{General scheme of RL for bioprocess control. The agent (controller) interacts with the environment (process) by selecting actions (inputs) based on the \textit{observed} system state. Upon sensing, the agent receives rewards and iteratively updates its policy to maximize the expected value of a user-defined return function (objective function).}
\label{fig:rl_general_scheme}
\end{figure*}

Reinforcement learning (RL) based on policy gradients, the focus of this article, is an alternative machine-learning-based control strategy for bioprocesses (cf. e.g., \cite{petsagkourakis_reinforcement_2020, espinel-rios_enhancing_2024}). In this framework, an agent (or \textit{controller}) interacts with the environment (or \textit{process}) by taking actions (or \textit{inputs}) and receiving rewards upon the agent's observations (or \textit{sensing}). Through this iterative process, the agent learns a control policy that maximizes the expected value of a user-defined return function (or \textit{objective function}) (Fig. \ref{fig:rl_general_scheme}; cf. \cite{sutton_reinforcement_2018, dong_introduction_2020} for more details on RL). Since RL continuously learns by interacting with the environment, the policy’s performance is expected to improve over time, making it inherently adaptive. Additionally, RL policies account for \textit{future uncertainties}, incorporating feedback by design.

While RL is generally \textit{model-free}, mathematical models can serve as \textit{surrogate environments} for the systems to be controlled. This enables \textit{in silico} policy training in a safe and cost-effective environment before actual experimental implementation. This approach is particularly advantageous in biotechnological processes, where running experiments can be time-consuming and expensive. Furthermore, domain knowledge can be leveraged to incorporate uncertainty into the surrogate model, allowing for a comprehensive robustness evaluation.

In policy-gradient RL, the policy is directly parameterized, e.g., via deep neural networks, and its parameters are iteratively updated using gradient ascent \cite{petsagkourakis_reinforcement_2020, espinel-rios_enhancing_2024}. This approach guarantees convergence to at least a local optimum with respect to the \textit{real} policy function, and control actions can be sampled \textit{directly} from the policy. As a result, policy-gradient methods are well-suited for continuous action spaces, which is advantageous in bioprocess control as it increases the degrees of freedom available for input modulation.

Managing control tasks with multiple objectives, such as multi-setpoint and multi-trajectory tracking, is nontrivial in RL. Throughout this work, we refer to \textit{setpoint tracking} as the task of following a reference that remains \textit{constant}, whereas \textit{trajectory tracking} refers to following a reference that varies over time. Although quadratic cost functions are well-established in {(model-based)} optimal control for multi-reference tracking \cite{hilo_mpc}, they exhibit limitations when applied to analogous problems in {(model-free)} RL {due to the additive nature of individual reward gains}. The challenge of appropriately weighting different reward components often results in unstable or slow learning, and in some cases, prevents the agent from learning the task altogether, as demonstrated in the case study of this work.

{In other RL applications, e.g., stabilizing an overhead crane at a desired position, a reward that combines a conventional quadratic cost with its logarithmic form has been used to amplify gains near the target \cite{zhang_deep_2023}. Although effective for single-objective tasks, extending such weighted formulations to multiple references (as in microbial consortia) adds complexity because each reference would require its own tuned weight.}

To address these challenges, we previously introduced an alternative return function specifically tailored for RL implementations of multi-setpoint tracking \cite{espinel-rios_enhancing_2024}\footnote{Accepted at the 14th IFAC Symposium on Dynamics and Control of Process Systems, including Biosystems (DYCOPS 2025).}. Our approach incentivizes the \textit{simultaneous} satisfaction of multiple setpoints while ensuring that no single objective dominates the learning process. This is achieved through multiplicative reciprocal saturation functions, which significantly enhance learning stability and control performance by providing the agent with a clearer gradient toward improving the overall control task. In other words, if one setpoint improves while others remain suboptimal, the overall reward is \textit{penalized} as a result of the inherent multiplicative \textit{coupling} of rewards in the return function. {In contrast to conventional quadratic-cost-based return functions, the multiplicative saturation-based functions provide balanced learning without manual weighting of reward components associated with individual references.}

Here, we extend our previous work by:  
1) systematically evaluating the method on different setpoint combinations (beyond setpoints of equal value) and analyzing the impact of tunable parameters in the return function on the RL outcome;  
2) extending our analysis beyond multi-setpoint tracking to multi-trajectory tracking, a more challenging control task; and  
3) assessing the robustness of our proposed RL method by considering system uncertainty in both initial conditions and key model parameters.  In all test cases, we compare our approach against the benchmark quadratic-cost-based return function.

The remainder of this paper is structured as follows. Section \ref{sec:control_problem} introduces the formulation of the stochastic control problem, which serves as the foundation for the RL framework using policy gradients, described in Section \ref{sec:RL_methodology}. In Section \ref{sec:RL_return_function}, we present the return functions considered in this study, including our proposed saturation-based return function and the benchmark quadratic-cost-based return function. Recognizing the growing importance of consortium-based bioprocesses, we apply our method to a biotechnologically relevant case study in Section \ref{sec:bioprocess_case_study}, focusing on population-level control via cybergenetic growth modulation through optogenetics.

\section{General formulation of the stochastic control problem}
\label{sec:control_problem}
As a preface to our stochastic control problem, let us first consider a \textit{deterministic} system dynamics which can be described in \textit{discrete} form as:
\begin{equation}
    \bm{x}_{t+1} = \bm{f_x}(\bm{x}_t, \bm{u}_t), \quad \forall t \in \{0, 1, \dots, N_s-1\}, \label{eq:dynamics_det}
\end{equation}
where $ \bm{x}_t \in \mathbb{R}^{n_x} $ represents the state vector at time step $t$, $ \bm{u}_t \in \mathbb{R}^{n_u} $ denotes the control input vector at time step $t$, and $ \bm{f_x}: \mathbb{R}^{n_x} \times \mathbb{R}^{n_u} \to \mathbb{R}^{n_x} $ is the state transition function. We assume equidistant sampling intervals of length $ \Delta t $ between consecutive states. We consider stepwise constant control actions applied over $ N_s $ intervals, leading to a total of $ N_s+1 $ discrete states. The final discrete time step is denoted by the subscript $ N_s $, corresponding to a continuous-time value of $ t_f = N_s \Delta t $. The initial condition is given by $\bm{x}_0 \in \mathbb{R}^{n_x}$ at $t_0 = 0$. 

Many bioprocesses are subject to uncertainties, such as uncertain initial conditions, uncertain model parameters, stochastic gene expression, and process disturbances. These uncertainties are challenging to capture within a deterministic control framework. Therefore, within the context of RL, we consider the system dynamics in a \textit{probabilistic} manner. To achieve this, we reformulate the discretized system dynamics presented in Eq. \eqref{eq:dynamics_det} as a Markov decision process. Specifically, the state transition is governed by the probability distribution $\bm{x}_{t+1} \sim \mathrm{P}(\bm{x}_{t+1} \mid \bm{x}_t, \bm{u}_t)$, where $\mathrm{P}$ denotes the conditional probability distribution of the next state $\bm{x}_{t+1}$ given the current state $\bm{x}_t$ and control input $\bm{u}_t$. 

In that sense, we can approximate the stochastic behavior of the plant by modeling the state transition with a function influenced by \textit{random} disturbances $\bm{d}_t \in \mathbb{R}^{n_d}$:
\begin{equation}
    \bm{x}_{t+1} = \bm{f_s}(\bm{x}_t, \bm{u}_t, \bm{d}_t), \quad \forall t \in \{0, 1, \dots, N_s-1\}, \label{eq:dynamics_stoc}.
\end{equation}
Here, $\bm{f_s}: \mathbb{R}^{n_x} \times \mathbb{R}^{n_u} \times \mathbb{R}^{n_d} \to \mathbb{R}^{n_x}$ is the stochastic state transition function that maps the current state $\bm{x}_t$, control input $\bm{u}_t$, and disturbances $\bm{d}_t$ to the next state $\bm{x}_{t+1}$. These random disturbances can be sampled from various sources, such as probabilistic distributions of model parameters, initial conditions, and process disturbances, which may be modeled using, e.g., Gaussian noise.

Furthermore, within the context of RL, we aim to maximize the \textit{expectation} $\mathbb{E}[\cdot]$ of a \textit{stochastic} objective function $J_s(\bm{\tau})$, referred to as the \textit{return function}: 
\begin{equation} 
    \max_{\pi(\cdot)} \mathbb{E}_{\bm{\tau}} \left[ J_s(\bm{\tau}) \right], 
    \label{eq:rl_problem} 
\end{equation} 
where $\pi(\cdot)$ denotes the \textit{stochastic} policy, which maps the observed system state $\bm{s}_t \in \mathbb{R}^{n_s}$ to a probability distribution over actions. In other words, the agent samples actions at each time step given the current system observation $\bm{s}_t \in \mathbb{R}^{n_s}$ and parameters $\bm{\theta} \in \mathbb{R}^{n_\theta}$ which \textit{shape} the probability function:
\begin{equation}
\bm{u}_{t} \sim \pi(\bm{u}_t \mid \bm{s}_t, \bm{\theta}).
\end{equation}
In Section \ref{sec:RL_return_function}, we outline the specific return functions considered for multi-setpoint and multi-trajectory tracking problems. The expectation in Eq. \eqref{eq:rl_problem} is taken over a trajectory $\bm{\tau}$ generated under the policy $\pi(\cdot)$, consisting of a sequence of \textit{observed} states, actions, and rewards: 
\begin{equation} 
    \bm{\tau} = \{ (\bm{s}_0, \bm{u}_0, R_1, \bm{s}_1), (\bm{s}_1, \bm{u}_1, R_2, \bm{s}_2), \ldots, (\bm{s}_{N_s-1}, \bm{u}_{N_s-1}, R_{N_s}, \bm{s}_{N_s}) \}, 
\end{equation} 
where $R_{t+1} \in \mathbb{R}$ represents the system reward, quantifying the \textit{benefit gain} of taking action $\bm{u}_t$ given the observed state $\bm{s}_t$ at time $t$. Note that rewards are assigned only after actions have been executed and the system has transitioned to its next state.

In this work, we assume that the control policy is normally distributed with mean $\bm{m}_t \in \mathbb{R}^{n_u}$ and standard deviation $\bm{\sigma}_t \in \mathbb{R}^{n_u}$. Both $\bm{m}_t$ and  $\bm{\sigma}_t$ are modeled using deep neural networks $\bm{f_\mathrm{DNN}}: \mathbb{R}^{n_s} \times \mathbb{R}^{n_\Theta} \rightarrow \mathbb{R}^{n_u} \times \mathbb{R}^{n_u}$:
\begin{equation}
\bm{m}_t, \bm{\sigma}_t = \bm{f_\mathrm{DNN}} (\bm{s}_t, \bm{\Theta}),
\label{eq:mean_std_policy}
\end{equation}
parametrized by $\Theta \in \mathbb{R}^{n_\Theta}$. Thus, we define $\bm{\theta} := \bm{\Theta}$ for consistency in notation. The parameter vector $\bm{\theta}$ will be the main focus of the policy optimization in Section \ref{sec:RL_methodology}. 

Note the system \textit{observation} $\bm{s}_t$ in Eq. \eqref{eq:mean_std_policy} works as the \textit{feature space} in a machine-learning context, allowing flexibility in selecting relevant \textit{features} as the agent's observation to inform the agent's decision-making process. These features may include \textit{measured} dynamic states, previously applied inputs, and the current process time, among others. 

With these ideas in mind, and following the chain rule of probability, the conditional probability of $\bm{\tau}$ reads:
\begin{equation}
\mathrm{P}(\bm{\tau} \mid \bm{\theta}) = \mathrm{P}(\bm{x}_0) \cdot \prod_{t=0}^{N_s-1} \left[ \pi(\bm{u}_t \mid \bm{s}_t, \bm{\theta}) \cdot \mathrm{P}(\bm{x}_{t+1} \mid \bm{x}_t, \bm{u}_t) \right].
\label{eq:prob_dis_tau}
\end{equation}
Thus, the likelihood of a trajectory $\bm{\tau}$ is expressed as the product of the initial state probability, the stochastic policy, and the state transition probabilities.

\section{Reinforcement learning via policy gradients}
\label{sec:RL_methodology}
To determine the optimal input policy's parameters, we consider \textit{gradient ascent}:
\begin{equation}
\bm{\theta}_{m+1} = \bm{\theta}_m + \alpha \nabla_{\bm{\theta}} \mathbb{E}_{\bm{\tau}} \left[ J_s(\bm{\tau}) \right], \quad \forall m \in \{0, 1, \dots, N_m-2\}.
\label{eq:update_rule_general}
\end{equation}
Here, the subscript $m$ denotes an \textit{epoch}, i.e., an update step, while $\alpha \in \mathbb{R}$ is the learning rate or step size in the direction of the gradient ascent. Note that before the first update at $m=0$, the policy parameters are randomly initialized. The first policy update is denoted as $\theta_0$, and the process continues iteratively, leading to $N_m$ policies: $\theta_0, \theta_1, ..., \theta_{N_m-1}$.

To compute $\mathbb{E}_{\bm{\tau}} \left[ J_s(\bm{\tau}) \right]$, we consider the Policy Gradient Theorem \cite{NIPS1999_464d828b}. Therefore:
\begin{equation}
\nabla_{\bm{\theta}} \mathbb{E}_{\bm{\tau}} \left[ J_s(\bm{\tau}) \right] = \nabla_{\bm{\theta}} \int \mathrm{P}(\bm{\tau} \mid \bm{\theta}) \cdot J_s(\bm{\tau}) \, \mathrm{d}\bm{\tau} = \int \nabla_{\bm{\theta}} \mathrm{P}(\bm{\tau} \mid \bm{\theta}) \cdot J_s(\bm{\tau}) \, \mathrm{d}\bm{\tau} = \int \mathrm{P}(\bm{{\tau}} \mid \bm{\theta}) \cdot \nabla_{\bm{\theta}} \log \mathrm{P}(\bm{\tau} \mid \bm{\theta}) \cdot J_s(\bm{\tau}) \, \mathrm{d}\bm{\tau},
\label{eq:policy_gradient_theorem}
\end{equation}
which leads to:
\begin{equation}
\nabla_{\bm{\theta}} \mathbb{E}_{\bm{\tau}} \left[ J_s(\bm{\tau}) \right] = \mathbb{E}_{\bm{\tau}} [ J_s(\bm{\tau}) \cdot \nabla_{\bm{\theta}} \log \mathrm{P}(\bm{\tau}|\bm{\theta})].
\label{eq:sim_eq_gradient_exp}
\end{equation}

For convenience, we reformulate $\nabla_{\bm{\theta}} \log \mathrm{P}(\bm{\tau}|\bm{\theta})$ in Eq. \eqref{eq:sim_eq_gradient_exp}. First, we take the logarithm of Eq. \eqref{eq:prob_dis_tau} and use the property that the logarithm of a product is the sum of the individual logarithms. We then simplify it by removing the gradients of terms that do not depend on $\bm{\theta}$, allowing us to rewrite Eq. \eqref{eq:sim_eq_gradient_exp} as:
\begin{equation}
\nabla_{\bm{\theta}} \mathbb{E}_{\bm{\tau}} \left[ J_s(\bm{\tau}) \right] = \mathbb{E}_{\bm{\tau}} \left[ J_s(\bm{\tau}) \cdot \nabla_{\bm{\theta}} \left[\sum_{t=0}^{N_s-1} \log \pi(\bm{u}_t \mid \bm{s}_t, \bm{\theta}) \right] \right].
\label{eq:grad_exp}
\end{equation}

The \textit{intractable} expectation is approximated via Monte Carlo sampling. To improve stability, we normalize the return function by subtracting the {mean return $\Bar{J}_{s_m}$ and dividing by the standard deviation of the return $\sigma_{J_{s_m}}$ in the epoch}. A small \textit{machine epsilon} constant $\epsilon_\mathrm{mach}$ is used in the denominator to avoid division by zero. Thus, Eq. \eqref{eq:grad_exp} is reformulated as:
\begin{equation}
\nabla_{\bm{\theta}} \mathbb{E}_{\bm{\tau}} \left[ J_s(\bm{\tau}) \right] \approx \frac{1}{N_\mathrm{MC}} \sum_{k=1}^{N_\mathrm{MC}} \left[ \frac{J_s\left( \bm{\tau}^{(k)} \right) - \Bar{J}_{s_m}\left( \bm{\tau} \right)}{\sigma_{J_{s_m}} + \epsilon_\mathrm{mach}} \cdot \nabla_{\bm{\theta}} \left[ \sum_{t=0}^{N_s-1} \log \left( \pi(\bm{u}_t^{(k)} \mid \bm{s}_t^{(k)}, \bm{\theta}) \right) \right] \right],
\label{eq:grad_exp_2}
\end{equation}
where $N_\mathrm{MC}$ represents the number of sampled trajectories of $\bm{\tau}$ or \textit{episodes}. Each episode is indicated by the superscript $(\cdot)^{(k)}$. The difference $(J_s(\bm{\tau}^{(k)}) - \Bar{J}_{s_m}(\bm{\tau}))$ determines the relative contribution of \textit{each} trajectory’s gradient (cf. Eq. \eqref{eq:sim_eq_gradient_exp}) in the parameter update (cf. Eq. \eqref{eq:update_rule_general}). Since the gradient is computed from the log-probability of the trajectory, trajectories with higher-than-average returns $(J_s(\bm{\tau}^{(k)}) > \Bar{J}_{s_m}(\bm{\tau}))$ increase the probability that the agent selects the actions that led to those trajectories, this drives the gradient ascent process to refine the policy in that direction.

It should be noted that even if the system to be controlled behaves deterministically, allowing a stochastic policy \textit{by design}, where actions are sampled from probability distributions, can help the agent \textit{explore} a wider range of actions during the learning process. Over time, the policy may still converge to a deterministic behavior, i.e., distributions with negligible standard deviations, but maintaining stochasticity during training remains beneficial, e.g., in escaping local minima.
 
\section{Return functions for multi-setpoint and multi-trajectory tracking}
\label{sec:RL_return_function}
Below, we outline the two return functions we consider in this study for tracking multiple setpoints and trajectories: the quadratic cost-based function and the multiplicative reciprocal saturation function. 

\subsection{Quadratic-cost-based function}
\label{subsub:quadratic_cost_function}
This is formulated as the \textit{inverse} (i.e., negated) quadratic cost commonly used in optimal control. This transformation aligns with the \textit{maximization} objective of the \textit{expected} \textit{reward-based} return function (cf. Eq. \eqref{eq:rl_problem}), which differs from the \textit{minimization} objective of a \textit{cost function} typically used in optimal control:
\begin{subequations}
    \begin{align}
        J_{s} &:= -\left[ \sum_{t=1}^{N_s-1} l_{s,q}(\bm{x}_t) + e_{s,q}(\bm{x}_{N_s}) \right], \label{eq:quadratic_cost_stoc_1} \\
        l_{s,q}(\bm{x}_t) &:= \| \bm{x}_t - \bm{x}_{t}^* \|^2_{\mathbf{Q}}, \quad \forall t \in \{1, ..., N_s-1\}, \label{eq:quadratic_cost_stoc_2} \\
        e_{s,q}(\bm{x}_{N_s}) &:= \| \bm{x}_{N_s} - \bm{x}_{N_s}^* \|^2_{\mathbf{Q_T}}, \label{eq:quadratic_cost_stoc_3}
    \end{align}
\end{subequations}
where $ l_{s,q}: \mathbb{R}^{n_x} \to \mathbb{R} $ and  
$ e_{s,q}: \mathbb{R}^{n_x} \to \mathbb{R} $ are the quadratic-cost \textit{stage} and \textit{terminal} rewards, respectively. Furthermore, $ \bm{x}_{t}^* \in \mathbb{R}^{n_x} $ and $ \bm{x}_{N_s}^* \in \mathbb{R}^{n_x} $ are state reference vectors. It is important to remark that the key distinction between multi-setpoint and multi-trajectory tracking lies in the reference: \textit{setpoint tracking uses a constant reference, while trajectory tracking follows a time-varying reference}. The weight matrices $ \mathbf{Q} \in \mathbb{R}^{n_x \times n_x} $ and $ \mathbf{Q_T} \in \mathbb{R}^{n_x \times n_x} $ determine the importance of tracking errors in the stage cost and terminal rewards, respectively. Note that $\| \bm{a} \|^2_\mathbf{A} := \bm{a}^\tran \mathbf{A} \bm{a}$ denotes the squared norm of a vector $\bm{a}$ weighted by the matrix $\mathbf{A}$. In this formulation, states that are not tracked are assigned zero stage and terminal weights.

Here, the maximum achievable return is zero, corresponding to perfect tracking $\bm{x}_t = \bm{x}_t^*$. Since the return function follows a Markov decision process, it starts accumulating \textit{rewards} only after the first action is taken, i.e., from discrete time subscript $t=1$.

\begin{figure*}[htb!]
    \centering
    % First row
    \begin{subfigure}{\textwidth}
        \centering
        \begin{tabular}{m{0.4\textwidth} m{0.05\textwidth} m{0.4\textwidth}}
            \begin{subfigure}{\linewidth}
                \centering
                \includegraphics[width=\linewidth, trim=0pt 10pt 1pt 10pt, clip]{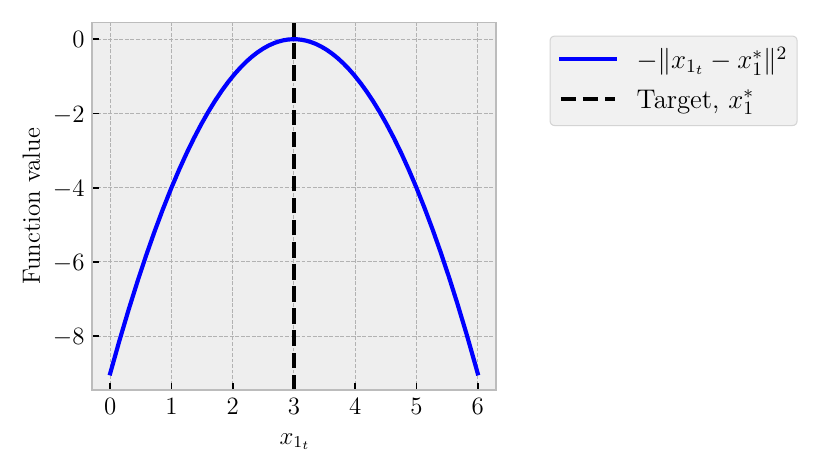}
            \end{subfigure} 
            & 
            \centering
            \raisebox{1.8\height}{\textbf{\scalebox{3}{%
                \tikz[baseline] {
                \draw[line width=2pt] (-0.8ex, 0) -- (0.8ex, 0);  % Horizontal bar
                \draw[line width=2pt] (0, -0.8ex) -- (0, 0.8ex);  % Vertical bar
            }}}} 
            & 
            \begin{subfigure}{\linewidth}
                \centering
                \includegraphics[width=\linewidth, trim=0pt 10pt 0pt 10pt, clip]{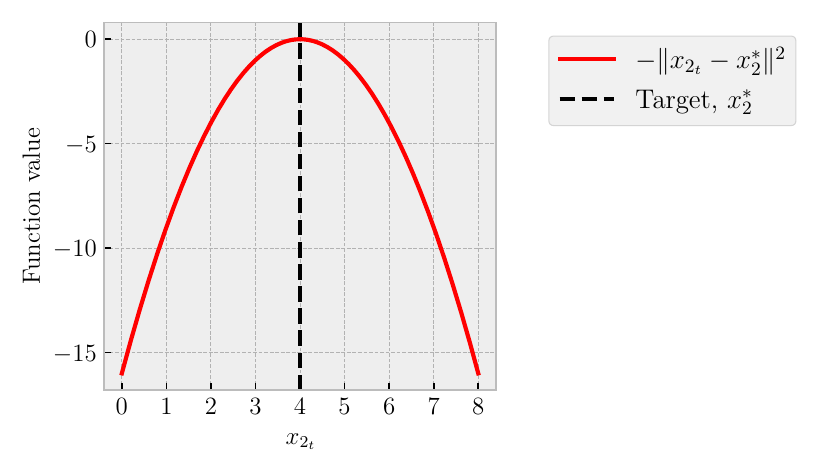}
            \end{subfigure} 
        \end{tabular}
        \vspace{-0.5em} % Adjust spacing between figure and subcaption
        \centering
        \subcaption{\textbf{Negated quadratic cost functions in summation}}
        \label{fig:addition}
    \end{subfigure}
    \\[1em]
    % Second row
    \begin{subfigure}{\textwidth}
        \centering
        \begin{tabular}{m{0.4\textwidth} m{0.05\textwidth} m{0.4\textwidth}}
            \begin{subfigure}{\linewidth}
                \centering
                \includegraphics[width=\linewidth, trim=0pt 10pt 1pt 10pt, clip]
                {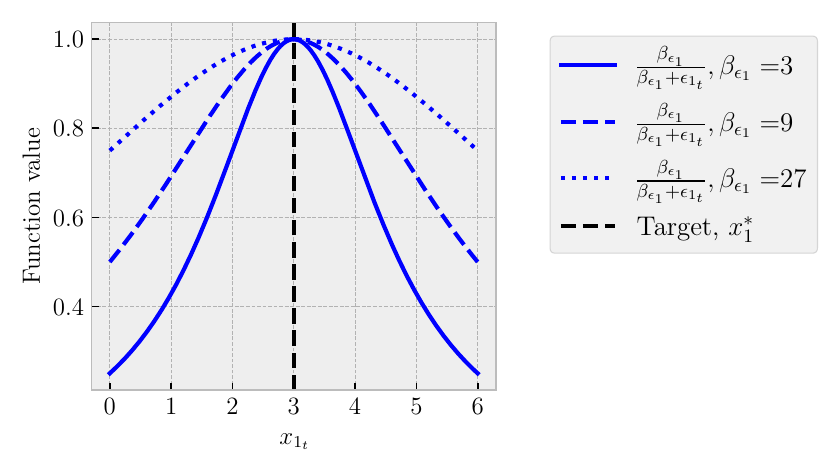}
                %left, bottom, right, top
            \end{subfigure} 
            & 
            \centering
            \raisebox{0.0\height}{\textbf{\scalebox{3}{\tikz\draw[line width=2pt] (0,0) -- (1ex,1ex) (0,1ex) -- (1ex,0);}}}  
            & 
            \begin{subfigure}{\linewidth}
                \centering
                \includegraphics[width=\linewidth, trim=0pt 10pt 0pt 10pt, clip]{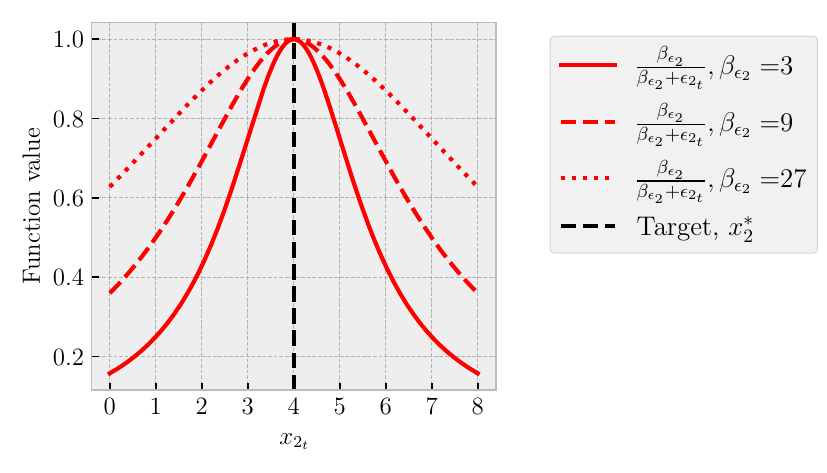}
            \end{subfigure} 
        \end{tabular}
        \vspace{-0.5em} % Adjust spacing between figure and subcaption
        \centering
        \subcaption{\textbf{Reciprocal saturation functions in multiplication}}
        \label{fig:multiplication}
    \end{subfigure}
    \caption{Illustration of the return functions analyzed in this work for two arbitrary tracked states $x_1$ and $x_2$ at a given sampling time $t$. A) A negated quadratic cost function (\textit{the benchmark in this study}), where the tracking squared errors of individual states are summed. In this example, the maximum return value at this sampling time is $0+0=0$. B) A multiplicative saturation function (\textit{our proposed approach}), where tracking errors are incorporated into the product of reciprocal saturation functions. This approach \textit{scales down} or \textit{penalizes} the return if one state deviates significantly, promoting coordinated learning of the control task. In this example, the maximum return value at this sampling time is $1\times1=1$. {Although only two tracked states are plotted, the same logic applies to any number of tracked states.}}
    \label{fig:comparison_return_functions}
\end{figure*}

To better understand the \textit{qualitative} behavior of the quadratic-cost-based function in RL, consider a scenario with two states to be tracked. Since the reward contributions of both tracked states are independent and appear as additive terms, as illustrated in Fig. \ref{fig:comparison_return_functions}-A, the agent may become biased toward improving only one objective or may fail to learn in a stable and smooth manner, as the objectives can shift between different references over epochs. This occurs because there is no mechanism guiding the learning process \textit{toward simultaneously meeting both references}, ultimately limiting control performance in the overall system.

\subsection{Multiplicative reciprocal saturation function}
We propose a return function based on reciprocal \textit{saturation} functions to address the challenges associated with quadratic-cost-based return functions. This function couples the overall rewards to the requirement of accurately tracking \textit{all} setpoints and trajectories. Mathematically, this is formulated as follows:
\begin{subequations}
    \begin{align}
        J_{s} &:= \sum_{t=1}^{N_s-1} l_{s,c}(\bm{x}_t) + e_{s,c}(\bm{x}_{N_s}), \label{eq:multiplicative_return_1} \\
        l_{s,c}(\bm{x}_t) &= w_t \left[ \alpha_{\text{max}} \prod_{i \in \mathcal{X}_{\text{track}}} \frac{\beta_{\epsilon_i}}{\beta_{\epsilon_i} + \epsilon_{i_t}} \right], \quad \forall t \in \{1, ..., N_s-1\}, \label{eq:multiplicative_return_2} \\
        e_{s,c}(\bm{x}_{N_s}) &= w_{N_s} \left[\alpha_{\text{max}} \prod_{i \in \mathcal{X}_{\text{track}}} \frac{\beta_{\epsilon_{i}}}{\beta_{\epsilon_{i}} + \epsilon_{i_{N_s}}} \right],\label{eq:multiplicative_return_3} \\
        \epsilon_{i_t} &= \| x_{i_t} - x_{i}^* \|^2,\,  \epsilon_{i_{N_s}} = \| x_{i_{N_s}} - x_{i}^* \|^2 \label{eq:multiplicative_return_4}.
    \end{align}
\end{subequations}
The notation of the stage and terminal rewards in Eqs. \eqref{eq:multiplicative_return_1}-\eqref{eq:multiplicative_return_4} follows that of Eqs. \eqref{eq:quadratic_cost_stoc_1}-\eqref{eq:quadratic_cost_stoc_3}, with the subscript $(\cdot)_{s,c}$ indicating the \textit{coupling} nature of the return function. {$\epsilon_{i_t}$ and $\epsilon_{i_{N_s}}$ represent the tracking error in the form of the squared deviation of the tracked state $i$ from its reference value at a stage time $t$ and at a terminal time $N_s$, respectively}. In addition, $w_t$ and $w_{N_s}$ are weighting parameters that balance the contributions of the different reward components throughout the sampling times. The parameter $\alpha_{\text{max}}$ determines the maximum achievable reward at a given time step when all tracking errors approach zero. The parameter $\beta_{\epsilon_{i}}$ determines the smoothness and steepness of the reciprocal saturation function, as illustrated in Fig. \ref{fig:comparison_return_functions}-B. This can strongly influence the learning dynamics, as will be demonstrated in the case study. This constant can be interpreted as the \textit{error half‐saturation constant}, and determines the error level at which the saturation function drops to half its maximum value. Finally, $\mathcal{X}_{\text{track}} \subseteq \{1, \dots, n_x\}$ represents the set of tracked states in multi-setpoint and multi-trajectory problems, with $x_{i}^* \in \mathbb{R}$ being the reference for a state $x_{i}$ in $\mathcal{X}_{\text{track}}$. The number of tracked states is given by the cardinality $|\mathcal{X}_{\text{track}}|$. {Note that, unlike the quadratic cost where errors are summed directly inside the norm, each state error ($\epsilon_{i_t}, \epsilon_{i_{N_s}}$) in the saturation-based approach is handled individually and then combined multiplicatively in $l_{s,c}(\bm{x}_t)$ and $e_{s,c}(\bm{x}_{N_s})$.}

To better understand the qualitative behavior of our proposed return function in RL, consider a scenario with two states to be tracked. Since the reward contributions of both tracked states are now coupled through the multiplication of reciprocal saturation functions with respect to the tracking error (cf. Fig.~\ref{fig:comparison_return_functions}-B), any deviation from a single reference significantly reduces or \textit{cancels} the overall reward. In other words, the simultaneous satisfaction of all references is required for maximum reward accumulation. This guides the agent to reduce the tracking error in all states, rather than focusing on only a subset, thereby providing better properties for stable learning and overall control efficiency, as will be demonstrated with the case study.

{\textit{Remark}. The design of the return function in
Eqs.~\eqref{eq:multiplicative_return_1}–\eqref{eq:multiplicative_return_4}
is inspired by the multi-substrate Monod equation, widely used in
bioprocess engineering to express growth rate as a function of several
limiting nutrients \cite{liu_how_2017}. However, we consider \textit{reciprocal} saturation terms, meaning the reward varies \textit{inversely} with the tracking error. Without loss of generality, let us consider a stage time $t$. Starting from the usual
hyperbolic saturation:
\begin{equation}
\frac{\epsilon_{i_t}}{\epsilon_{i_t}+\beta_{\epsilon_i}}=\frac{1}{1+\beta_{\epsilon_i}/\epsilon_{i_t}},
\end{equation}
we take the \textit{reciprocal} of the ratio $(\beta_{\epsilon_i}/\epsilon_{i_t})$ to obtain:
\begin{equation}
\frac{1}{1+\epsilon_{i_t}/\beta_{\epsilon_i}}=\;\frac{\beta_{\epsilon_i}}{\beta_{\epsilon_i} + \epsilon_{i_t}},
\end{equation}
which is the general form used in the formulation of our saturation-based return function.}

{\textit{Remark}. Without loss of generality, let us consider a stage reward $l_{s,c}(\bm{x}_t)$ and $w_t=1$. Here, the fraction $\tfrac{\beta_{\epsilon_i}}{\beta_{\epsilon_i} + \epsilon_{i_t}}$ acts as an efficiency factor between 0 and 1. Thus, for a finite error, the stage reward satisfies:
\begin{equation}
  0 < l_{s,c}(\bm{x}_t)
    = \alpha_{\text{max}}
      \prod_{i\in\mathcal X_{\text{track}}}
      \frac{\beta_{\epsilon_i}}{\beta_{\epsilon_i}+\epsilon_{i_t}}
    \;\leq\; \alpha_{\text{max}},
  \qquad
  l_{s,c}(\bm{x}_t) = \alpha_{\text{max}}
  \;\Longleftrightarrow\;
  \epsilon_{i_t}=0\;\forall i .
\end{equation}
Hence the maximum stage reward can be reached \textit{only} when all tracking errors are zero, thereby guiding the agent to satisfy every reference tracking objective simultaneously.}

\section{Cybergenetic case study: two-member consortium of \textit{E. coli} with optogenetic control of growth}
\label{sec:bioprocess_case_study}
To demonstrate the efficiency and robustness of our novel return function for RL implementations involving multi-setpoint and multi-trajectory tracking, we consider a two-member consortium of \textit{Escherichia coli} growing in a chemostat. Similar to \cite{espinel-rios_enhancing_2024}, we assume that both strains consume glucose as a carbon source and do not have any engineered co-dependency interactions. Furthermore, we assume that the cells are engineered for external optogenetic control of auxotrophic behavior. Specifically, \textit{E. coli} 1 is auxotrophic for lysine upon deletion of \textit{lysA} (diaminopimelate decarboxylase), while \textit{E. coli} 2 is auxotrophic for leucine upon deletion of \textit{leuA} (2-isopropylmalate synthase). The expression of both \textit{lysA} and \textit{leuA} is regulated by blue and red light intensity, respectively, allowing external optogenetic control of growth. We assume that the PBLind-v1 system \cite{jayaraman_blue_2016} enables gene expression control via blue light, while the pREDawn-DsRed system \cite{multamaki_optogenetic_2022} achieves similar control using red light. Additionally, we assume that amino acid induction does not result in excretion, as the systems are designed to accumulate amino acids only up to normal physiological levels, sufficient for full growth restoration. 

\begin{figure*} [htb!]
\begin{center}
\includegraphics[scale=0.35]{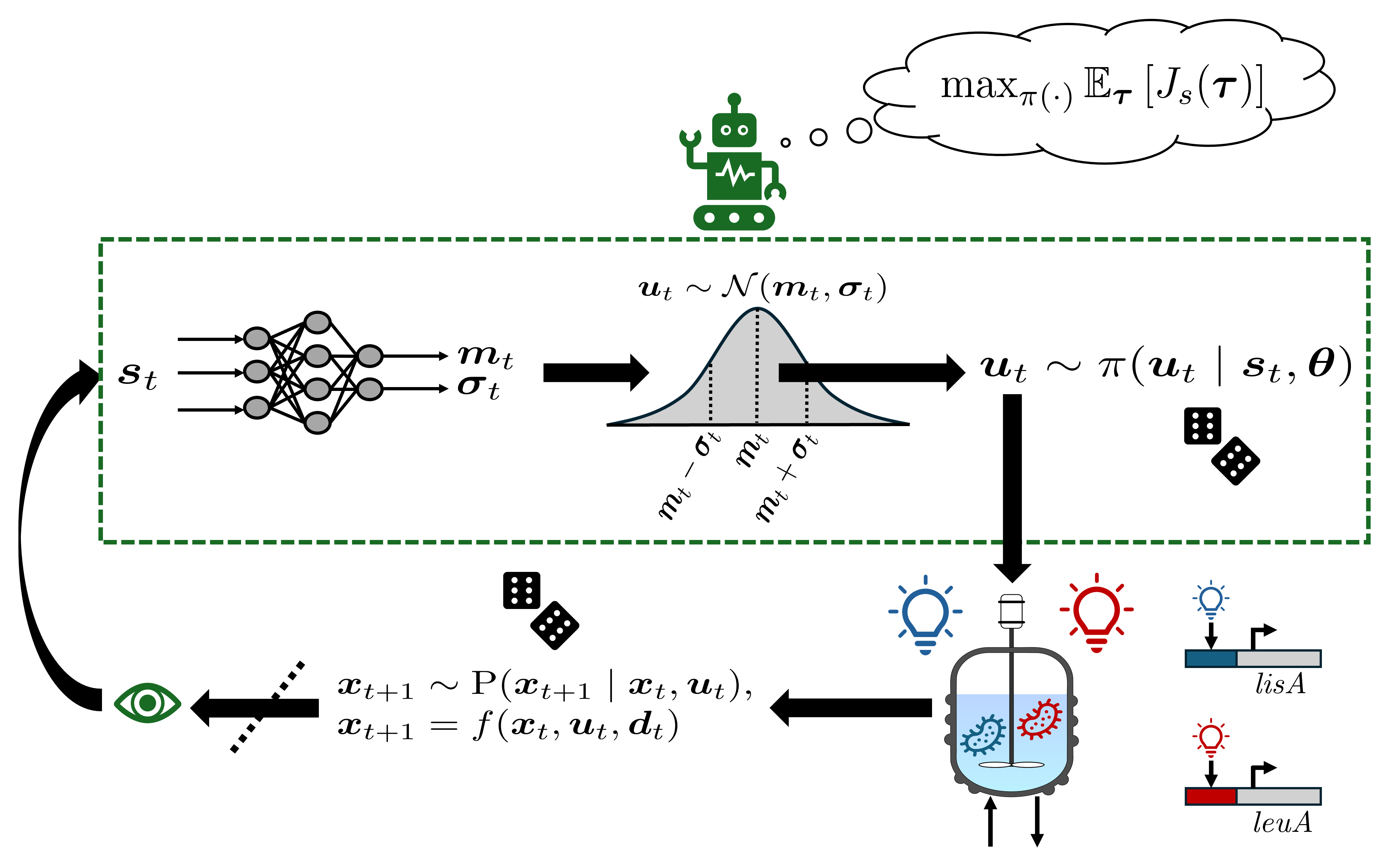}
\end{center}
\caption{Overview of the computational case study. Cybergenetic control of microbial growth via optogenetic regulation of amino-acid-based auxotrophy. Blue light modulates \textit{lysA} (diaminopimelate decarboxylase), which controls the production of essential amino acid lysine, while red light modulates \textit{leuA} (2-isopropylmalate synthase), which controls the production of essential amino acid leucine. The RL agent aims to optimally track multiple setpoints and dynamic trajectories by maximizing the user-defined return function (cf. Sections \ref{sec:control_problem}-\ref{sec:RL_return_function} for details on the methodology and notation).}
\label{fig:methodology_scheme}
\end{figure*}

\subsection{Dynamic model of the cybergenetic system}
\label{sec:bioprocess_model}
For our computational experiments, we consider the following system dynamics in the chemostat:
\begin{subequations}
\begin{align}
    &\odv{g}{t} = - q_{g_1} b_1 - q_{g_2} b_2 + (g_{\text{in}} - g) d_l, \label{eq:s_ode}\\
    &\odv{b_i}{t} = (\mu_i - d_l) b_i, \quad \forall i\in\{1,2\}, \label{eq:b_i_ode}\\
    &\odv{a_i}{t} = q_{a_i} - (d_{a_i} + \mu_i) a_i, \quad \forall i\in\{1,2\} \label{eq:e_i_ode},
\end{align}
\end{subequations}
where $g \in \mathbb{R}$ represents the glucose concentration; the shared substrate. The biomass concentrations of \textit{E. coli} 1 and \textit{E. coli} 2 are denoted by $b_1 \in \mathbb{R}$ and $b_2 \in \mathbb{R}$, respectively.  Therefore, in the case study, $\mathcal{X}_{\text{track}}:= \{b_1, b_2\}$. Similarly, the intracellular concentrations of the amino acids lysine and leucine are denoted by $a_1 \in \mathbb{R}$ and $a_2 \in \mathbb{R}$, respectively. We consider two constant operational parameters, $d_l$ and $g_{\text{in}}$, which represent the constant dilution rate and the inflow substrate concentration, respectively. The amino acid degradation rate is represented by $d_{a_i}$.

The kinetic functions follow Monod-type kinetics for growth and substrate consumption, while amino acid production is described using Hill-type kinetics, lumping both optogenetic transcription and translation:
\begin{subequations}
\begin{align}
    &\mu_i = \mu_{\text{max}_i} \left( \frac{g}{g + k_{g_i}} \right) \left( \frac{ f_{c} a_i}{ f_{c} a_i + k_{a_i}} \right), \quad \forall i\in \{1,2\}, \label{eq:mu_i}\\
    &q_{g,i} = Y_{g/b_i} \mu_i, \quad \forall i\in \{1,2\}, \label{eq:q_s_i}\\
    &q_{a,i} = q_{a_{\text{max}_i}} \left( \frac{I_i^{n_i}}{I_i^{n_i} + k_{I_i}^{n_i}} \right), \quad \forall i\in \{1,2\} \label{eq:q_e_i}.
\end{align}
\end{subequations}
Here, $f_{c}$ is an appropriate conversion factor. In addition, for \textit{E. coli} strain $i$, $I_i$ represents the corresponding optogenetic light control input, and $Y_{g/b_i}$ is the yield of substrate on biomass. The parameters $k_{g_i}$, $k_{a_i}$, and $k_{I_i}$ are saturation constants, while $\mu_{\text{max}_i}$ and $q_{a_{\text{max}_i}}$ denote the maximum growth and amino acid production rates, respectively.  The nominal parameter values and initial conditions used in this study are listed in Table \ref{tab:parameters}.

\begin{table}[htb!]
    \begin{center}
    \begin{threeparttable}
    \caption{Nominal model parameters and initial conditions used in the computational experiments.}
    \begin{tabular}{l c c c}
        \hline
        \textbf{Item} & \textbf{Value} & \textbf{Unit} & \textbf{Ref.}\\
        \hline
        % Model parameters
        $\mu_{\max_1}, \mu_{\max_2}$        & 0.982                         & $\mathrm{h^{-1}}$             & Note 1  \\
        $k_{g_1}, k_{g_2}$                  & $2.964 \times 10^{-4}$        & $\mathrm{mmol/L}$             & \cite{SENN1994424}  \\
        $f_{c}$                             & 1100                          & $\mathrm{g/L}$                & Note 2  \\
        $k_{a_1}$                           & 1.7                           & $\mathrm{mmol/L}$             & Note 3  \\
        $k_{a_2}$                           & 0.182                         & $\mathrm{mmol/L}$             & Note 3  \\
        $Y_{g/b_1}, Y_{g/b_2}$              & 10.18                         & $\mathrm{mmol/g}$             & Note 1  \\
        $q_{a_{\max_1}}$                    & 0.337                         & $\mathrm{mmol/(g \cdot h)}$   & Note 4  \\
        $q_{a_{\max_2}}$                    & 0.036                         & $\mathrm{mmol/(g \cdot h)}$   & Note 4  \\
        $n_1$                               & 2                             & 1                             & \cite{jayaraman_blue_2016}  \\
        $k_{I_1}$                           & 1.052                         & $\mathrm{W/m^{2}}$            & \cite{jayaraman_blue_2016}  \\
        $n_2$                               & 4.865                         & 1                             & \cite{multamaki_optogenetic_2022}  \\
        $k_{I_2}$                           & 1.34                          & $\mathrm{\mu W/cm^{2}}$       & \cite{multamaki_optogenetic_2022}  \\
        $d_l$                               & 0.15                          & $\mathrm{h^{-1}}$             & This work  \\
        $g_{\text{in}}$                     & 200                           & $\mathrm{mmol/L}$             & This work  \\
        % Initial values
        $g(0)$                              & 1 (multi-setpoint); 50 (multi-trajectory)         & $\mathrm{{mmol/L}}$   & This work  \\
        $b_1(0)$                            & 0.005 (multi-setpoint); 3 (multi-trajectory)      & $\mathrm{g/L}$        & This work  \\
        $b_2(0)$                            & 0.005 (multi-setpoint); 4 (multi-trajectory)      & $\mathrm{g/L}$        & This work  \\
        $a_1(0)$                            & $1.545 \times 10^{-2}$ (multi-setpoint); $1.075\times 10^{-4}$ (multi-trajectory) & $\mathrm{mmol/g}$     & This work  \\
        $a_2(0)$                            & $1.655 \times 10^{-3}$ (multi-setpoint); $2.998\times 10^{-5}$ (multi-trajectory) & $\mathrm{mmol/g}$     & This work  \\
        \hline
        \label{tab:parameters}
    \end{tabular}    
    \end{threeparttable}
    \end{center}
    \footnotesize{\textbf{Note 1}. From flux balance analysis using the ECC2 model \cite{hadicke_ecolicore2_2017} under aerobic conditions and glucose as carbon source constrained by 10 $\mathrm{mmol/g_x/h}$ glucose uptake.}
    \footnotesize{\textbf{Note 2}. Conversion factor based on the total cell density \cite{milo_cell_2016}.}
    \footnotesize{\textbf{Note 3}. Assumed as biologically sound values.}
    \footnotesize{\textbf{Note 4}. Computed upon assuming steady state conditions of amino acid production, maximum rates, and saturation concentration of the amino acids $\sim 10k_{a_i}$ corrected by the cell density.}
    {\footnotesize{\textbf{Note 5.} For the multi-setpoint scenarios, the initial conditions correspond to low inoculum concentrations typically present at chemostat start-ups, and we assume an initial maximum-growth metabolic state. For the multi-trajectory scenarios, we start from the nominal setpoints already achieved in a preceding multi-setpoint run, representing situations in which an operator wishes to dynamically re-balance populations following a predefined path. This could be the case when re-tuning metabolic sub-modules (e.g., to favor a different product or intermediate) without restarting the process.}}
    \end{table}

\newpage
\subsection{Overview of control scenarios}
\label{sec:RL_evaluation}
We consider four control cases:
\begin{itemize}
    \item \textbf{Case 1: multi-setpoint tracking \textit{without} uncertainty}. To demonstrate the flexibility of our approach, we test the tracking of four different constant setpoint combinations in the co-culture. No system uncertainty is considered.
    \item \textbf{Case 2: multi-trajectory tracking \textit{without} uncertainty}. To show that our approach extends beyond constant setpoints, we test the tracking of two different dynamic trajectory combinations in the co-culture. No system uncertainty is considered.
    \item \textbf{Case 3: robust multi-setpoint tracking \textit{under} uncertainty}. To evaluate robustness, we test the tracking of a selected setpoint combination under uncertain initial conditions and model parameters.
    \item \textbf{Case 4: robust multi-trajectory tracking \textit{under} uncertainty}. To evaluate robustness, we test the tracking of a selected dynamic trajectory combination under uncertain initial conditions and model parameters.
\end{itemize}

In all control cases, we compare our novel return function (cf. Eqs. \eqref{eq:multiplicative_return_1}-\eqref{eq:multiplicative_return_4}) against the quadratic-cost-based benchmark function (cf. Eqs. \eqref{eq:quadratic_cost_stoc_1}-\eqref{eq:quadratic_cost_stoc_3}). Experiments of this type are denoted as qc. Furthermore, for ease of comparison, we normalize the return function in all trials, scaling each to the range $[0,1]$ based on its respective maximum return value.

\sloppy
\textit{Remark on the policy parametrization and global learning parameters}. 
{Based on previous work \cite{petsagkourakis_reinforcement_2020}}, we parametrized the policy distribution using a deep feedforward neural network with four hidden layers, each containing 20 nodes, and the LeakyReLU activation function with a negative slope of 0.1. We used two output linear layers (without activation functions): one predicts the means and the other predicts the standard deviations of the normally distributed probabilities for the two control inputs (blue and red light intensities, $\bm{u}:= [I_1, I_2]^\tran$). These outputs are then used to construct the policy distribution (cf. Eq. \eqref{eq:mean_std_policy}). That is, the input distributions for the blue and red light intensities share the same hidden layers but have separate output layers for their means and standard deviations. In addition, we used $N_\mathrm{MC} = 500$ episodes per epoch and a learning rate $\alpha = 0.001$, as in our previous work \cite{espinel-rios_enhancing_2024}. The agent's observation $\bm{s}_t$ consists of two past state/input pairs and a time embedding $t_n$, normalized to $t_n \in [-1,1]$. Assuming full state observability, the agent's observation is defined as: $s_t := [\bm{x}_{t-1}^\tran, \bm{u}_{t-2}^\tran, \bm{x}_{t}^\tran, \bm{u}_{t-1}^\tran, t_n]^\tran$, where \textit{empty} states and inputs are pre-filled with zero values until filled with the past time horizon. We considered 18 stepwise constant control actions per input, thus $N_s = 18$, of length $ \Delta t = 1\, \mathrm{h}$. The RL agent (controller) is trained in PyTorch \citep{pytorch}, and the environment (process) is simulated in CasADi \citep{Andersson2019}.

\subsubsection{Case 1: multi-setpoint tracking without uncertainty}
The four tested setpoints ($b_1^*, b_2^*$) were: (1,6), (2,5), (3,4), (3.5,3.5) in $\mathrm{g/L}$. Hereafter, we will omit the units of the references when clear from the context. For each combination, we evaluated different values of $\beta_{\epsilon_i}$ (cf. Eqs. \eqref{eq:multiplicative_return_2}-\eqref{eq:multiplicative_return_3}): $\beta_{\epsilon_i} = 3, 9, 27$. These are denoted as ${\beta\_3}$, $\mathrm{\beta\_9}$, and $\mathrm{\beta\_27}$, respectively. {This hyperparameter set was chosen as it covers different smoothness and steepness levels of the saturation-based return functions (cf. Fig. \ref{fig:comparison_return_functions}-B), and we wanted to elucidate which $\beta_{\epsilon_i}$ values would provide the best results overall across different multi-setpoint references.} We considered $N_m = 500$ epochs. 

Additionally, we tested different reward-weighting schemes in the saturation-based function:
\begin{itemize}
    \item Terminal-only reward (denoted as tr): terminal weight equal to 1, all other weights equal to 0.
    \item Equal-stage-terminal reward (denoted as 1\_sr\_1\_tr): all weights equal to 1.
    \item Slightly terminal-weighted reward (denoted as 1\_sr\_2\_tr): stage weights equal to 1, terminal weight equal to 2.
    \item More terminal-weighted reward (denoted as 1\_sr\_3\_tr): stage weights equal to 1, terminal weight equal to 3.
\end{itemize}
The motivation for increasing the terminal reward weight was to test whether the agent would improve in performance by it having a \textit{terminal target in mind}. To clarify the naming of the experiments, for example, $\mathrm{1\_sr\_1\_tr\_\beta\_27}$ refers to an experiment using an equal-stage-terminal reward scheme with $\beta_{\epsilon_i}=27$. Overall, we systematically tested 13 learning schemes per setpoint combination, thus in total 52 setpoint learning schemes. For computational efficiency, we implemented early stopping with a patience of 100, meaning the training process stops if no improvement in return function is observed for 100 consecutive epochs. To facilitate the comparison of the scenarios in control case 1, we use two metrics: the \textit{total} normalized average absolute error (NAAE) and the {normalized} area under the curve (NAUC) of the return function.

For an \textit{individual} tracked state $i$, $ \mathrm{NAAE}_i$ is defined as:
\begin{equation}
    \mathrm{NAAE}_{i} = \frac{1}{N_s} \sum_{t=1}^{N_s} \left| \frac{x_{i}^* - \Bar{x}_{i_t}}{x_{i}^*} \right|, \quad \forall i \in \mathcal{X}_{\text{track}}, \label{eq:naae_i_av}
\end{equation}
{where $\Bar{x}_{i_t}$ represents the mean value across episodes in the epoch yielding the highest mean return.}

The \textit{total} NAAE, considering all references, is then given by the average of the individual NAAE values:
\begin{equation}
    \mathrm{NAAE} = \frac{1}{|\mathcal{X}_{\text{track}}|} \sum_{i \in \mathcal{X}_{\text{track}}} \mathrm{NAAE}_{i}.
    \label{eq:total_naae}
\end{equation}
This metric quantifies tracking error, with lower $\mathrm{NAAE}$ values indicating better tracking performance. However, this metric alone does not account for learning efficiency across training epochs, including aspects such as stability and convergence.

Therefore, in addition, the NAUC is computed using the trapezoidal method to approximate the cumulative return over training epochs, effectively \textit{integrating} the return function across epochs. For a fair comparison across scenarios, we normalize it based on the number of trapezoids evaluated, i.e., intervals between epochs:
\begin{equation}
    \mathrm{NAUC} =  \frac{1}{N_m-1} \sum_{i=0}^{N_m-2} \frac{\Bar{J}_i^* + \Bar{J}_{i+1}^*}{2} \Delta m,
\end{equation}
where $\Delta m = 1$ is the distance between epochs. {$\Bar{J}^*$ is the normalized mean return function, scaled to the range $[0,1]$ based on the maximum value achieved, which enables direct comparison across scenarios with different return values.}
This metric captures both \textit{convergence speed} (how quickly the policy achieves high returns) and \textit{learning stability} (fewer oscillations between high and low returns). Thus, a higher $\mathrm{NAUC}$ value indicates faster convergence and more stable learning. However, this metric alone does not reflect the final accuracy of the learned control policy, as it focuses solely on the learning process. {For instance, there may be rapid \say{convergence} to a return value that does not necessarily perform well.}

\begin{figure*}[htb!]
    \centering
    \begin{subfigure}{0.4\textwidth}
      \captionsetup{justification=centering}
      \includegraphics[scale=0.5]{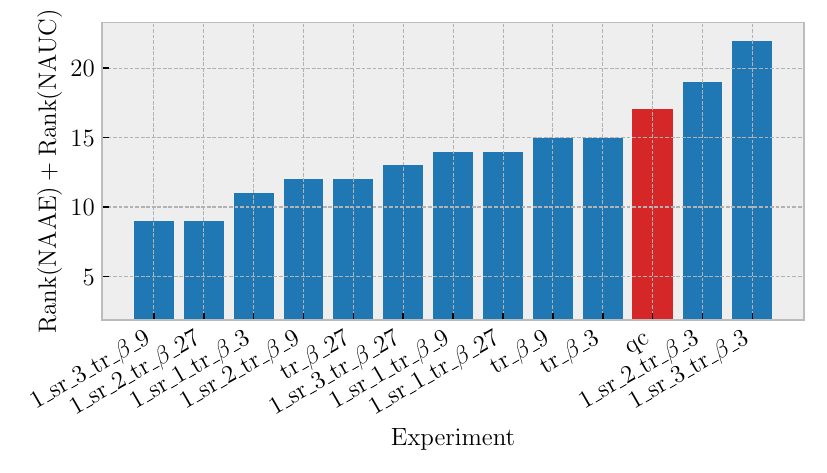}
      \subcaption[]{Setpoint: $b_1^* = 1, b_2^* = 6$}
    \end{subfigure}
    \hspace{1cm}
    \begin{subfigure}{0.4\textwidth}
      \captionsetup{justification=centering}
      \includegraphics[scale=0.5]{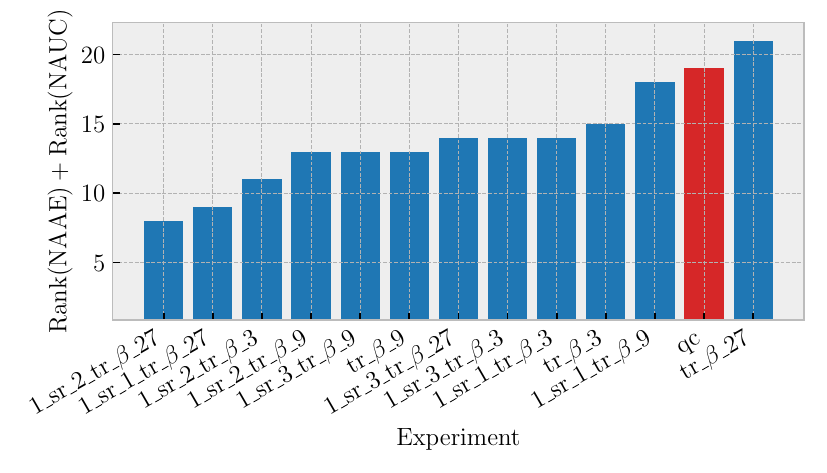}
      \subcaption[]{Setpoint: $b_1^* = 2, b_2^* = 5$}
    \end{subfigure}

    \begin{subfigure}{0.4\textwidth}
      \captionsetup{justification=centering}
      \includegraphics[scale=0.5]{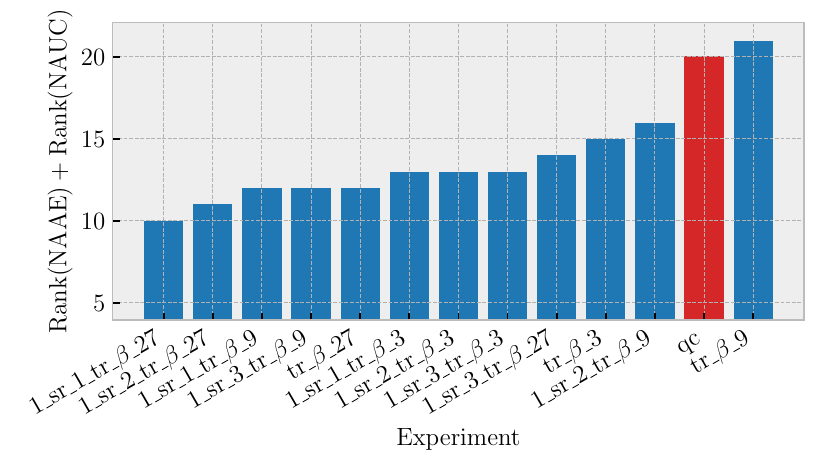}
      \subcaption[]{Setpoint: $b_1^* = 3, b_2^* = 4$}
    \end{subfigure}
    \hspace{1cm}
    \begin{subfigure}{0.4\textwidth}
      \captionsetup{justification=centering}
      \includegraphics[scale=0.5]{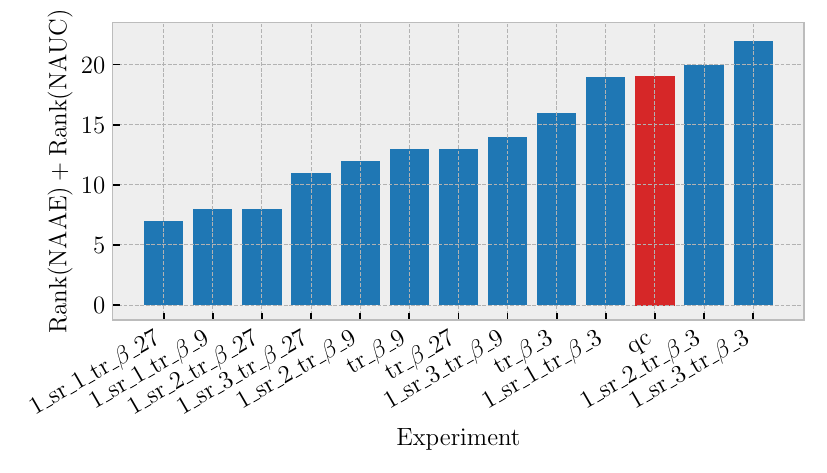}
      \subcaption[]{Setpoint: $b_1^* = 3.5, b_2^* = 3.5$}
    \end{subfigure}

    \caption{Bar plots systematically comparing the efficiency of different return-function configurations in RL control case 1 (multi-setpoint tracking without uncertainty) across various setpoint combinations of biomass populations ($b_1^*$ and $b_2^*$). The ranking of computational experiments is based on the combined ranks of both total $\mathrm{NAAE}$ and $\mathrm{NAUC}$. The benchmark quadratic-cost-based return function is highlighted in red.}
    \label{fig:bar_plots_rank}
\end{figure*}

With this in mind, we rank the total $\mathrm{NAAE}$ in ascending order and $\mathrm{NAUC}$ in descending order, {which allows for a fast preliminary exploration of control performance. We defined the best-performing control scenario as} the one that minimizes $\text{Rank}(\mathrm{NAAE}) + \text{Rank}(\mathrm{NAUC})$, with both ranks equally weighted for simplicity. {This composite metric is intended for hyperparameter screening; users may adopt a different weighting or formulation to suit specific requirements.} The ranking of return-function configurations for the tested setpoints in control case 1 is presented in Fig. \ref{fig:bar_plots_rank}. Regardless of the specific setpoint combination, the proposed reciprocal saturation-based return functions outperformed the benchmark quadratic-cost-based counterpart, the latter consistently ranking among the lowest-performing configurations. This was expected, given the ability of our proposed return function to incentivize the simultaneous satisfaction of references (in this case, setpoints), as discussed in Section \ref{sec:RL_return_function}. In addition, it is worth noting that the best-performing scenarios involved a combination of both stage and terminal rewards in the saturation-based return function, whereas using only the terminal reward led to overall poor performance, sometimes even worse than the benchmark. This was expected, as combining stage and terminal rewards provides the agent with a more comprehensive understanding of the process{; in other words, the return is computed from trajectories covering the process at all sampling instances}. Furthermore, we observed that the best-performing scenarios were associated with $\beta_{\epsilon_i}$ values of 27 and 9 in the saturation-based function, corresponding to the \textit{smoother} shapes of the functions (cf. Fig. \ref{fig:comparison_return_functions}-B). Intuitively, this can be attributed to the fact that smoother return functions produce less aggressive gradients in the gradient ascent update rule (cf. Eq. \eqref{eq:update_rule_general}), resulting in more stable learning dynamics and gradual parameter updates.

\FloatBarrier
\clearpage 
\begin{figure*}[h!]
    \centering
    %%%%%% REWARDS %%%%%% 
    \makebox[0.48\textwidth][c]{\textbf{Setpoint}: $b_1^* = 1, b_2^* = 6$, \textbf{no uncertainty}}
    \makebox[0.48\textwidth][c]{\textbf{Setpoint:} $b_1^* = 2, b_2^* = 5$, \textbf{no uncertainty}}\\
    \makebox[0.24\textwidth][c]{\textbf{Exp.}: 1\_sr\_3\_tr\_$\beta$\_9}
    \makebox[0.24\textwidth][c]{\textbf{Exp.}: qc}
    \makebox[0.24\textwidth][c]{\textbf{Exp.}: 1\_sr\_2\_tr\_$\beta$\_27}
    \makebox[0.24\textwidth][c]{\textbf{Exp.}: qc}\\
    \begin{subfigure}{0.24\textwidth}
      \captionsetup{justification=centering}
      \includegraphics[scale=0.4]{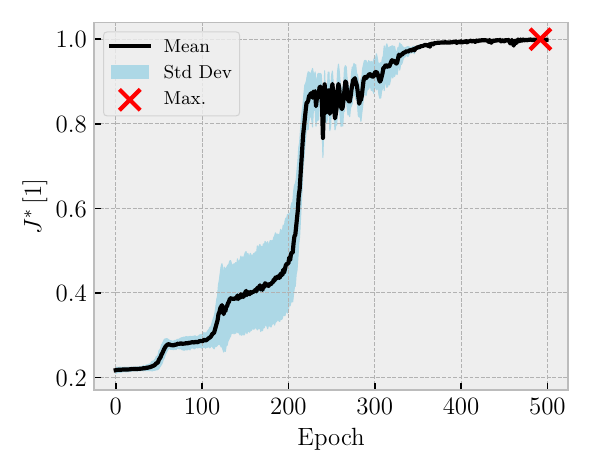}
      \subcaption[]{Reward}
    \end{subfigure}
    \begin{subfigure}{0.24\textwidth}
      \captionsetup{justification=centering}
      \includegraphics[scale=0.4]{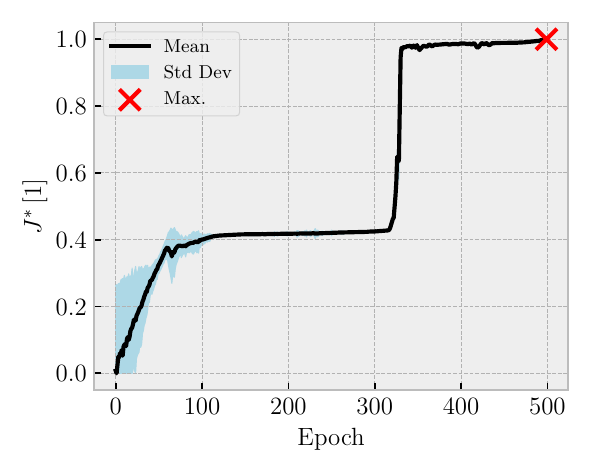}
      \subcaption[]{Reward}
    \end{subfigure}
    \vline
    \begin{subfigure}{0.24\textwidth}
      \captionsetup{justification=centering}
      \includegraphics[scale=0.4]{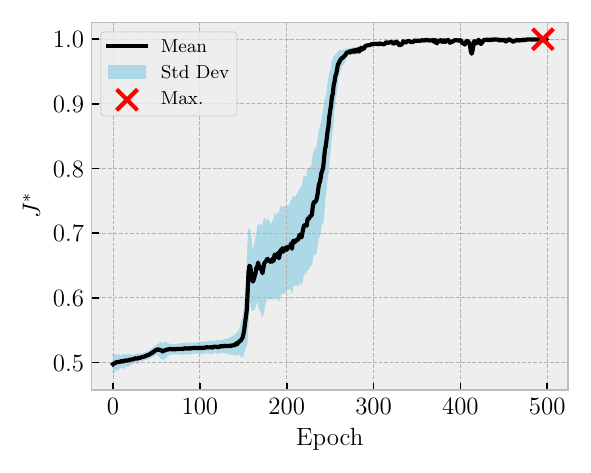}
      \subcaption[]{Reward}
    \end{subfigure}
    \begin{subfigure}{0.24\textwidth}
      \captionsetup{justification=centering}
      \includegraphics[scale=0.4]{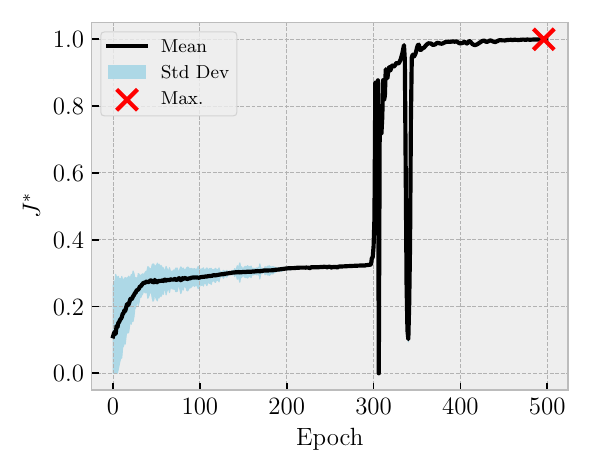}
      \subcaption[]{Reward}
    \end{subfigure}
    %%%%%% BIOMASS %%%%%% 
    \begin{subfigure}{0.24\textwidth}
      \captionsetup{justification=centering}
      \includegraphics[scale=0.4]{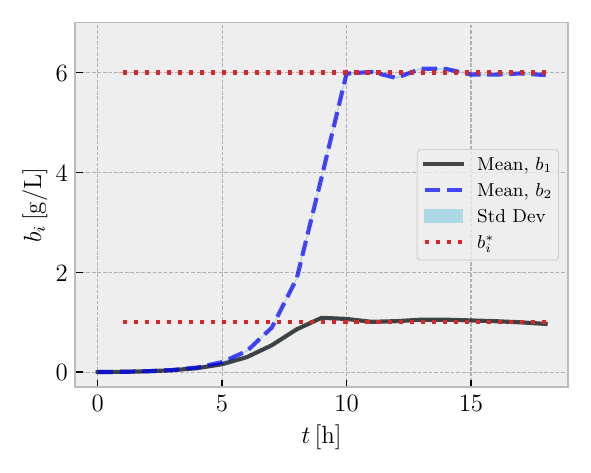}
      \subcaption[]{Biomass}
    \end{subfigure}
    \begin{subfigure}{0.24\textwidth}
      \captionsetup{justification=centering}
      \includegraphics[scale=0.4]{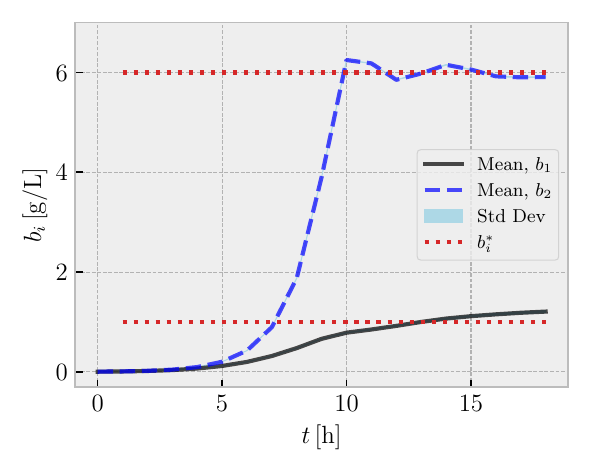}
      \subcaption[]{Biomass}
    \end{subfigure}
    \vline
    \begin{subfigure}{0.24\textwidth}
      \captionsetup{justification=centering}
      \includegraphics[scale=0.4]{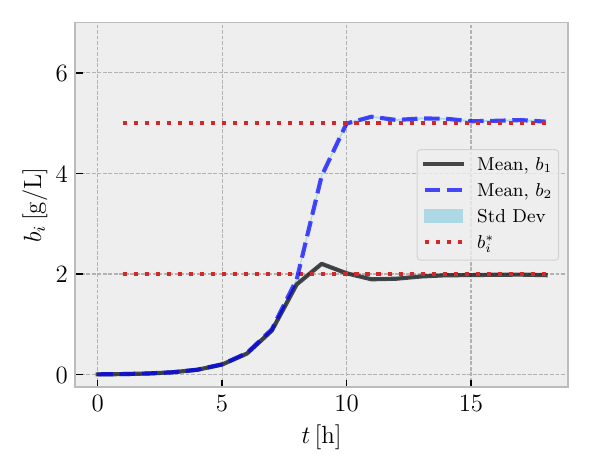}
      \subcaption[]{Biomass}
    \end{subfigure}
    \begin{subfigure}{0.24\textwidth}
      \captionsetup{justification=centering}
      \includegraphics[scale=0.4]{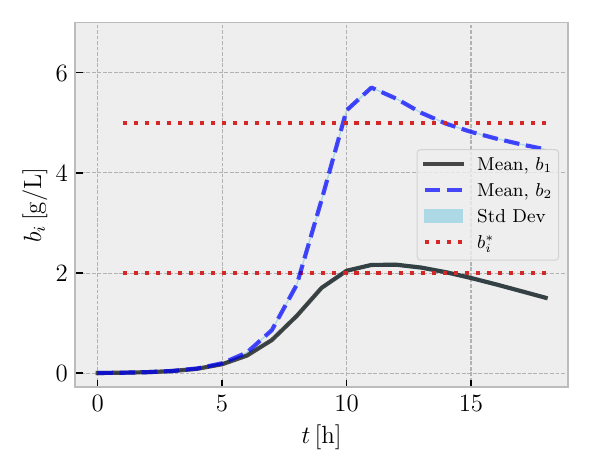}
      \subcaption[]{Biomass}
    \end{subfigure}
    %%%%%% GROWTH RATE %%%%%% 
    \begin{subfigure}{0.24\textwidth}
      \captionsetup{justification=centering}
      \includegraphics[scale=0.4]{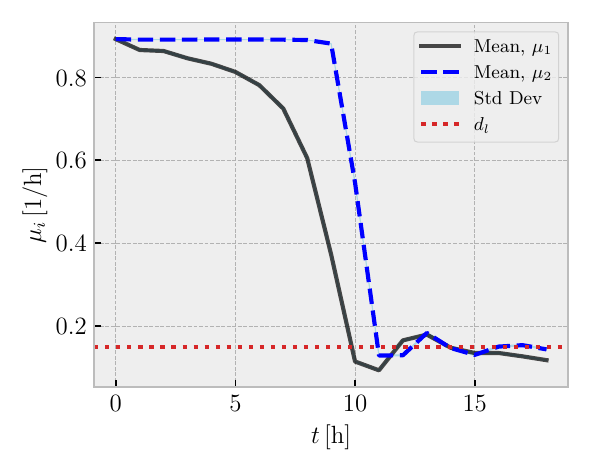}
      \subcaption[]{Growth rates}
    \end{subfigure}
    \begin{subfigure}{0.24\textwidth}
      \captionsetup{justification=centering}
      \includegraphics[scale=0.4]{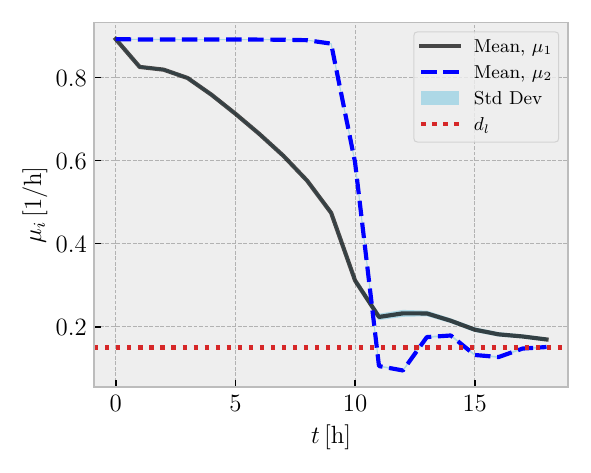}
      \subcaption[]{Growth rates}
    \end{subfigure}
    \vline
    \begin{subfigure}{0.24\textwidth}
      \captionsetup{justification=centering}
      \includegraphics[scale=0.4]{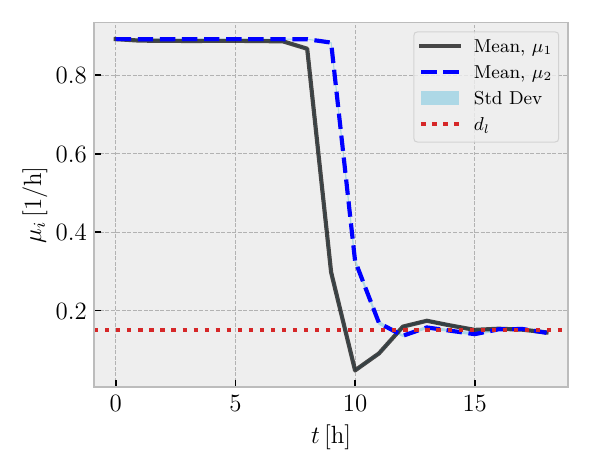}
      \subcaption[]{Growth rates}
    \end{subfigure}
    \begin{subfigure}{0.24\textwidth}
      \captionsetup{justification=centering}
      \includegraphics[scale=0.4]{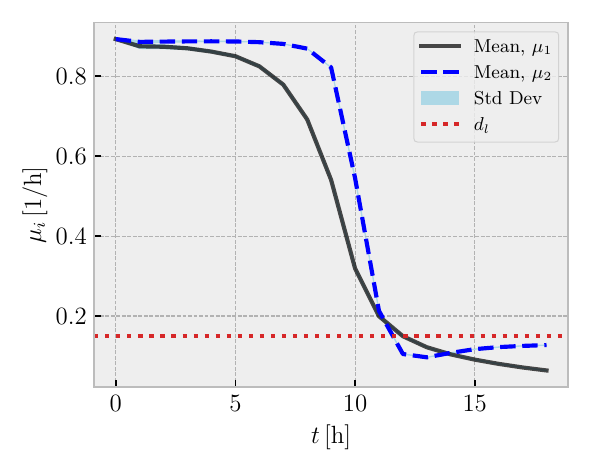}
      \subcaption[]{Growth rates}
    \end{subfigure}
    %%%%%% INPUT 1 %%%%%% 
    \begin{subfigure}{0.24\textwidth}
      \captionsetup{justification=centering}
      \includegraphics[scale=0.4]{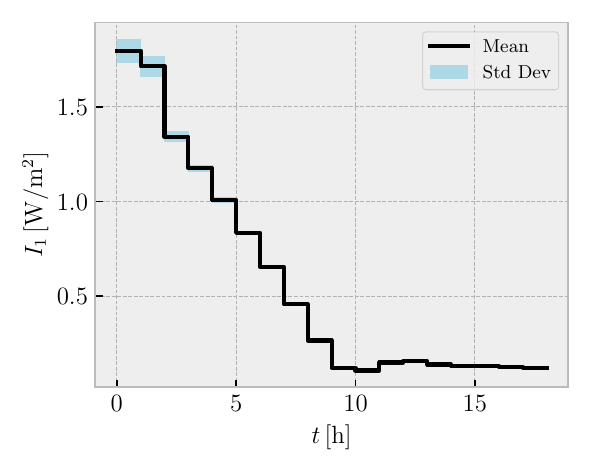}
      \subcaption[]{Input 1}
    \end{subfigure}
    \begin{subfigure}{0.24\textwidth}
      \captionsetup{justification=centering}
      \includegraphics[scale=0.4]{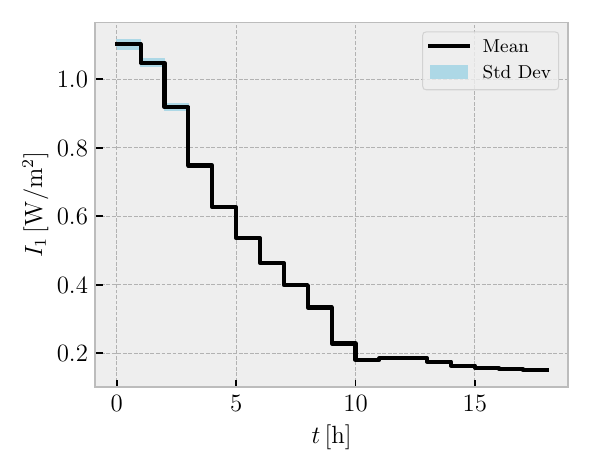}
      \subcaption[]{Input 1}
    \end{subfigure}
    \vline
    \begin{subfigure}{0.24\textwidth}
      \captionsetup{justification=centering}
      \includegraphics[scale=0.4]{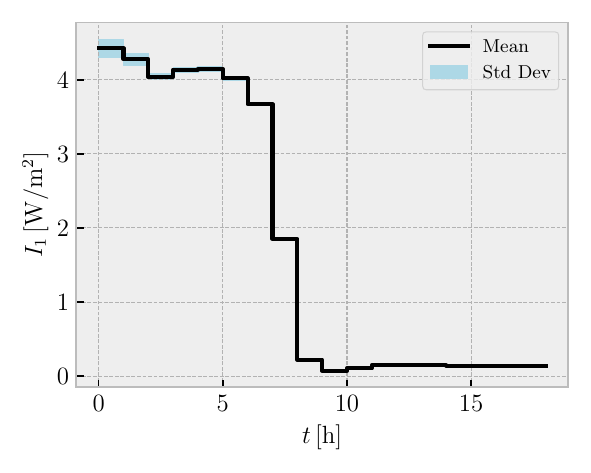}
      \subcaption[]{Input 1}
    \end{subfigure}
    \begin{subfigure}{0.24\textwidth}
      \captionsetup{justification=centering}
      \includegraphics[scale=0.4]{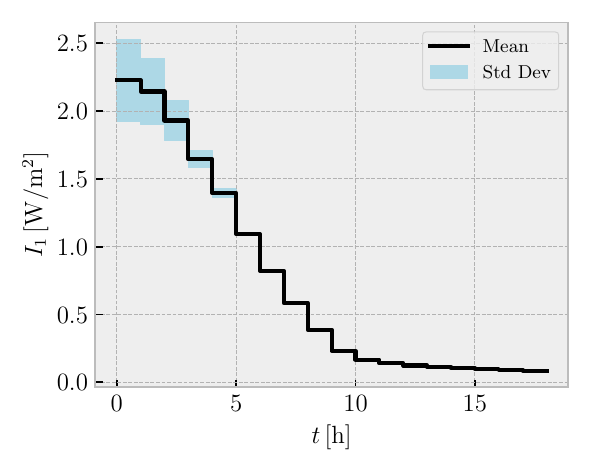}
      \subcaption[]{Input 1}
    \end{subfigure}
    %%%%%% INPUT 2 %%%%%% 
    \begin{subfigure}{0.24\textwidth}
      \captionsetup{justification=centering}
      \includegraphics[scale=0.4]{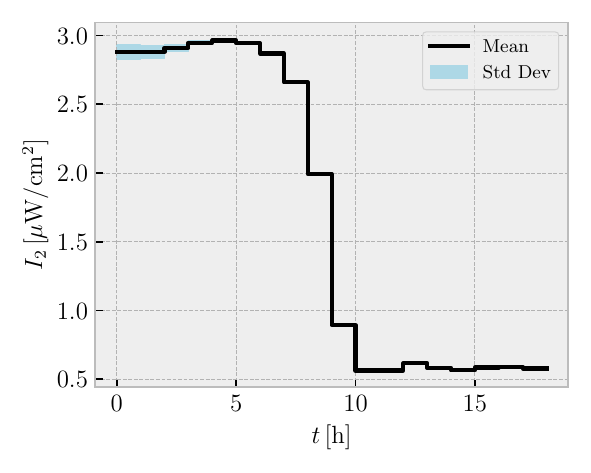}
      \subcaption[]{Input 2}
    \end{subfigure}
    \begin{subfigure}{0.24\textwidth}
      \captionsetup{justification=centering}
      \includegraphics[scale=0.4]{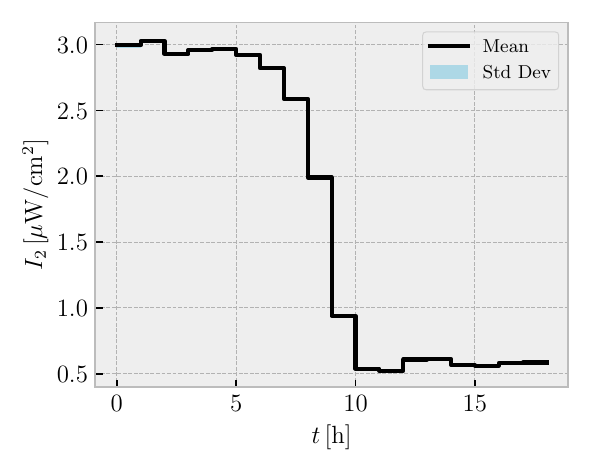}
      \subcaption[]{Input 2}
    \end{subfigure}
    \vline
    \begin{subfigure}{0.24\textwidth}
      \captionsetup{justification=centering}
      \includegraphics[scale=0.4]{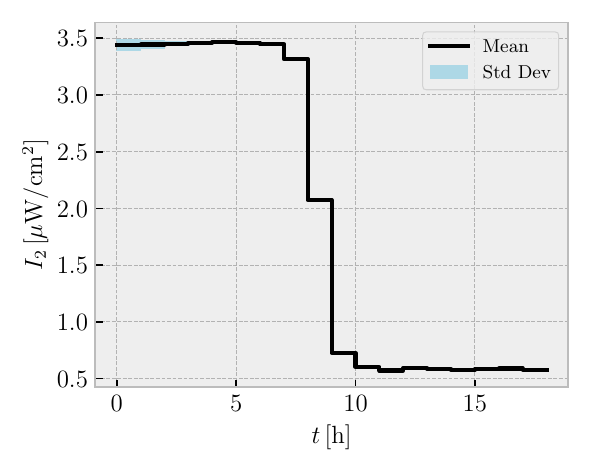}
      \subcaption[]{Input 2}
    \end{subfigure}
    \begin{subfigure}{0.24\textwidth}
      \captionsetup{justification=centering}
      \includegraphics[scale=0.4]{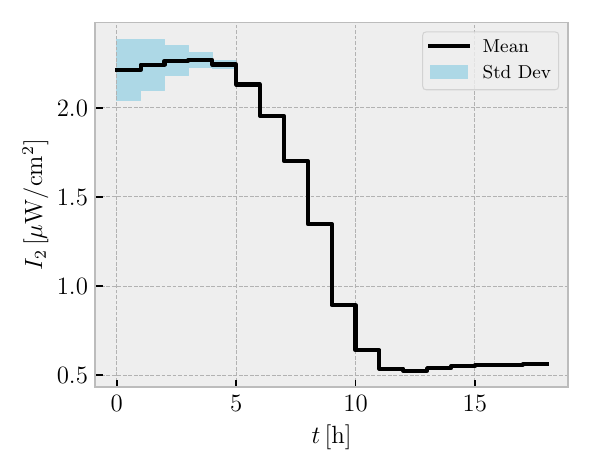}
      \subcaption[]{Input 2}
    \end{subfigure}
    \caption{Results for control case 1 (multi-setpoint tracking \textit{without} uncertainty) for setpoint combinations ($b_1^*, b_2^*$): $(1,6)$ and $(2,5)$. The \textit{normalized} return function $J^*$, scaled to the range $[0,1]$ based on the maximum value achieved, is plotted over all epochs until early stopping occurred or the maximum number of epochs was reached. Dynamic plots for biomass concentrations, growth rates, and applied inputs correspond to the epoch with the maximum mean return function value (red mark in the plot of the return function). The dotted red lines in the biomass plots represent the target setpoints, while the dotted red line in the plots of the growth rate represents the bioreactor's dilution rate. The blue shaded area indicates the standard deviation.}
    \label{fig:control_case_1_plots_part1}
\end{figure*}
\FloatBarrier
\clearpage 

\FloatBarrier
\clearpage 
\begin{figure*}[h!]
    \centering
    %%%%%% REWARDS %%%%%% 
    \makebox[0.48\textwidth][c]{\textbf{Setpoint}: $b_1^* = 3, b_2^* = 4$, \textbf{no uncertainty}}
    \makebox[0.48\textwidth][c]{\textbf{Setpoint:} $b_1^* = 3.5, b_2^* = 3.5$, \textbf{no uncertainty}}\\
    \makebox[0.24\textwidth][c]{\textbf{Exp.}: 1\_sr\_1\_tr\_$\beta$\_27}
    \makebox[0.24\textwidth][c]{\textbf{Exp.}: qc}
    \makebox[0.24\textwidth][c]{\textbf{Exp.}: 1\_sr\_1\_tr\_$\beta$\_27}
    \makebox[0.24\textwidth][c]{\textbf{Exp.}: qc}\\
    \begin{subfigure}{0.24\textwidth}
      \captionsetup{justification=centering}
      \includegraphics[scale=0.4]{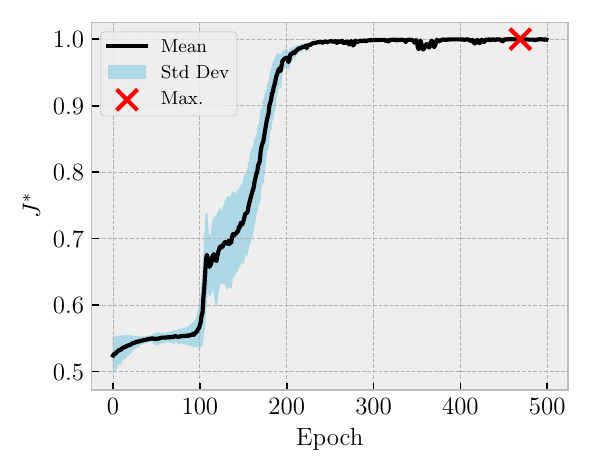}
      \subcaption[]{Reward}
    \end{subfigure}
    \begin{subfigure}{0.24\textwidth}
      \captionsetup{justification=centering}
      \includegraphics[scale=0.4]{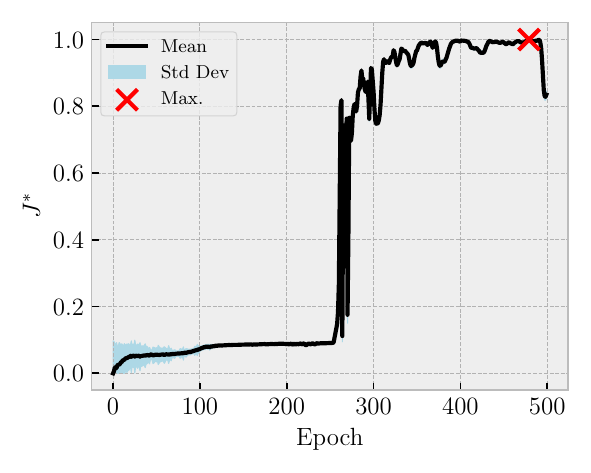}
      \subcaption[]{Reward}
    \end{subfigure}
    \vline
    \begin{subfigure}{0.24\textwidth}
      \captionsetup{justification=centering}
      \includegraphics[scale=0.4]{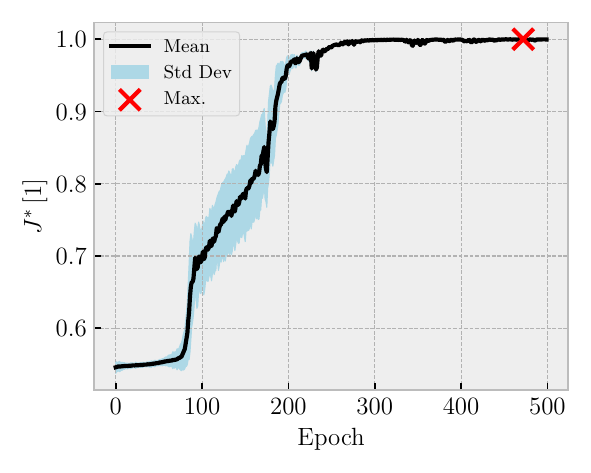}
      \subcaption[]{Reward}
    \end{subfigure}
    \begin{subfigure}{0.24\textwidth}
      \captionsetup{justification=centering}
      \includegraphics[scale=0.4]{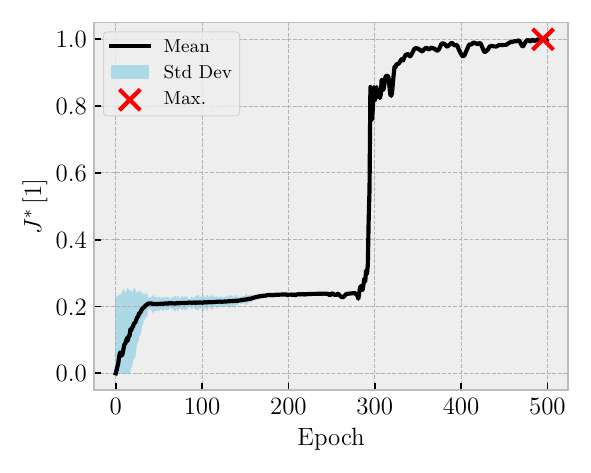}
      \subcaption[]{Reward}
    \end{subfigure}
    %%%%%% BIOMASS %%%%%% 
    \begin{subfigure}{0.24\textwidth}
      \captionsetup{justification=centering}
      \includegraphics[scale=0.4]{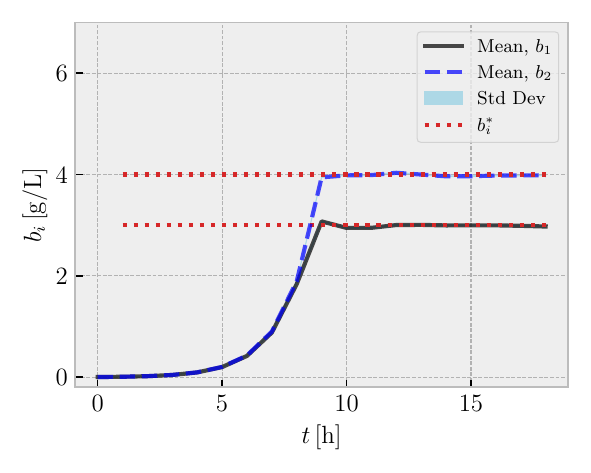}
      \subcaption[]{Biomass}
    \end{subfigure}
    \begin{subfigure}{0.24\textwidth}
      \captionsetup{justification=centering}
      \includegraphics[scale=0.4]{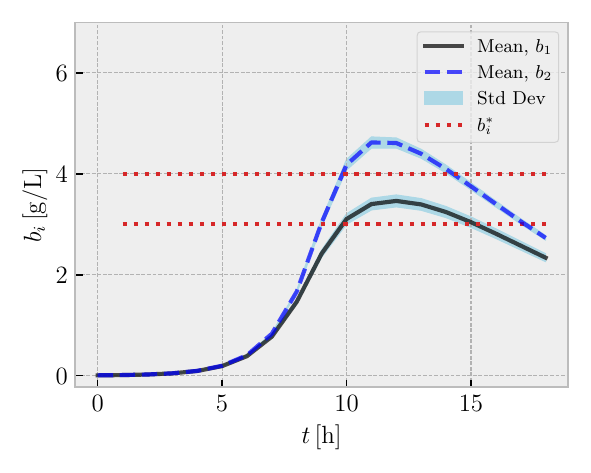}
      \subcaption[]{Biomass}
    \end{subfigure}
    \vline
    \begin{subfigure}{0.24\textwidth}
      \captionsetup{justification=centering}
      \includegraphics[scale=0.4]{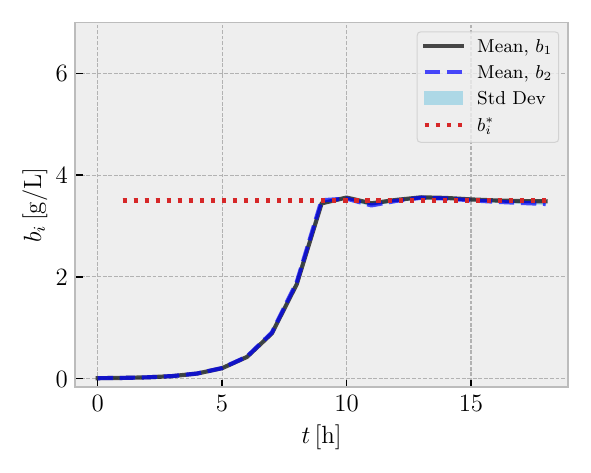}
      \subcaption[]{Biomass}
    \end{subfigure}
    \begin{subfigure}{0.24\textwidth}
      \captionsetup{justification=centering}
      \includegraphics[scale=0.4]{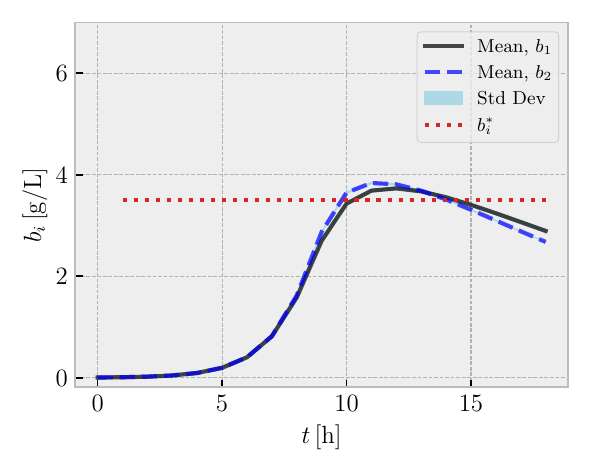}
      \subcaption[]{Biomass}
    \end{subfigure}
    %%%%%% GROWTH RATE %%%%%% 
    \begin{subfigure}{0.24\textwidth}
      \captionsetup{justification=centering}
      \includegraphics[scale=0.4]{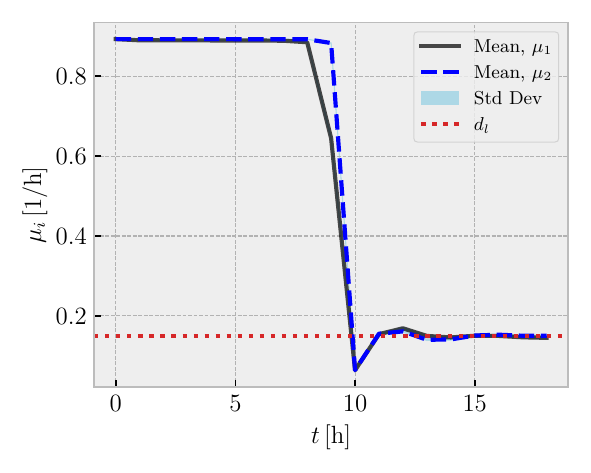}
      \subcaption[]{Growth rates}
    \end{subfigure}
    \begin{subfigure}{0.24\textwidth}
      \captionsetup{justification=centering}
      \includegraphics[scale=0.4]{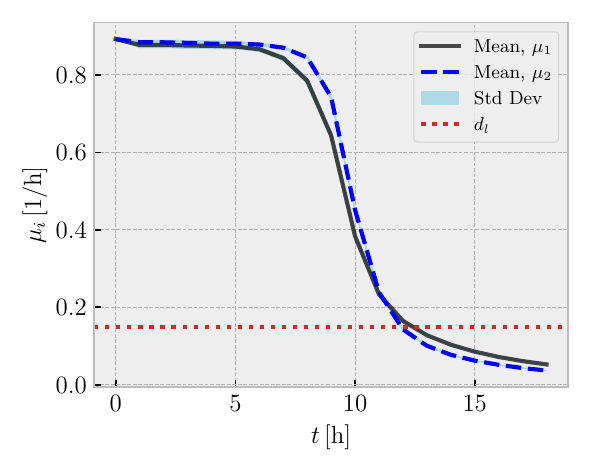}
      \subcaption[]{Growth rates}
    \end{subfigure}
    \vline
    \begin{subfigure}{0.24\textwidth}
      \captionsetup{justification=centering}
      \includegraphics[scale=0.4]{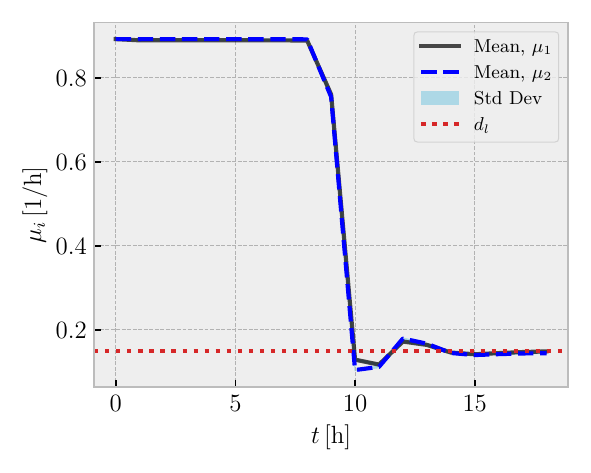}
      \subcaption[]{Growth rates}
    \end{subfigure}
    \begin{subfigure}{0.24\textwidth}
      \captionsetup{justification=centering}
      \includegraphics[scale=0.4]{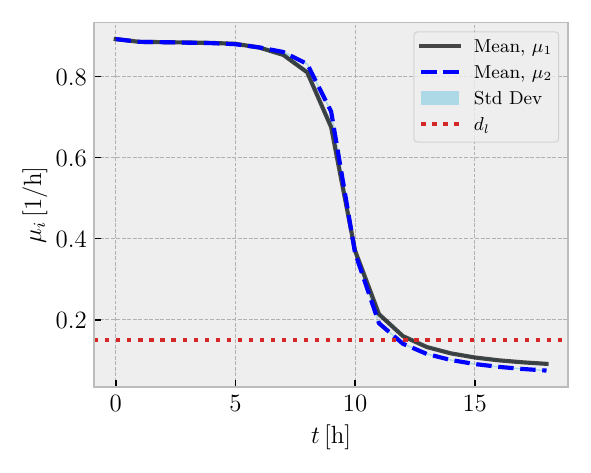}
      \subcaption[]{Growth rates}
    \end{subfigure}
    %%%%%% INPUT 1 %%%%%% 
    \begin{subfigure}{0.24\textwidth}
      \captionsetup{justification=centering}
      \includegraphics[scale=0.4]{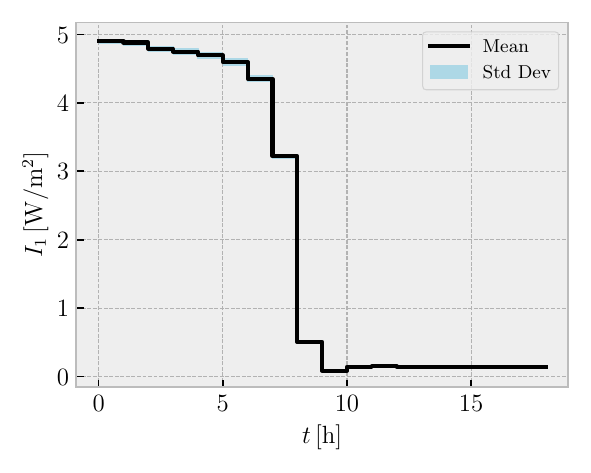}
      \subcaption[]{Input 1}
    \end{subfigure}
    \begin{subfigure}{0.24\textwidth}
      \captionsetup{justification=centering}
      \includegraphics[scale=0.4]{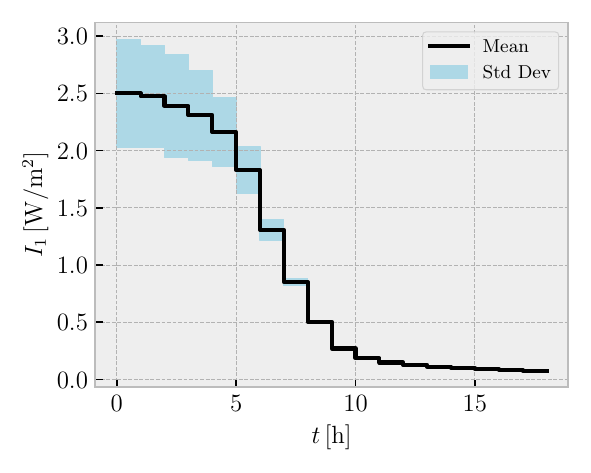}
      \subcaption[]{Input 1}
    \end{subfigure}
    \vline
    \begin{subfigure}{0.24\textwidth}
      \captionsetup{justification=centering}
      \includegraphics[scale=0.4]{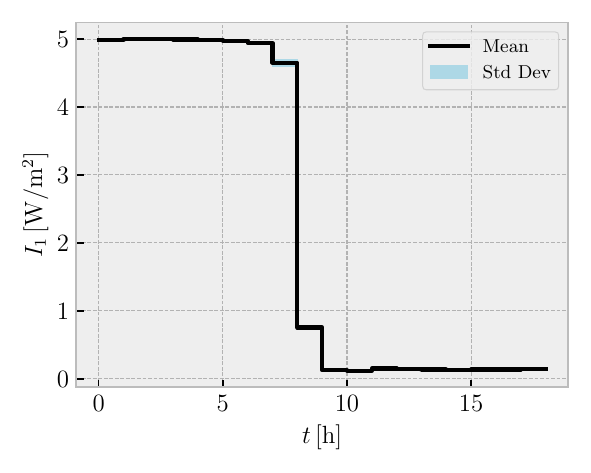}
      \subcaption[]{Input 1}
    \end{subfigure}
    \begin{subfigure}{0.24\textwidth}
      \captionsetup{justification=centering}
      \includegraphics[scale=0.4]{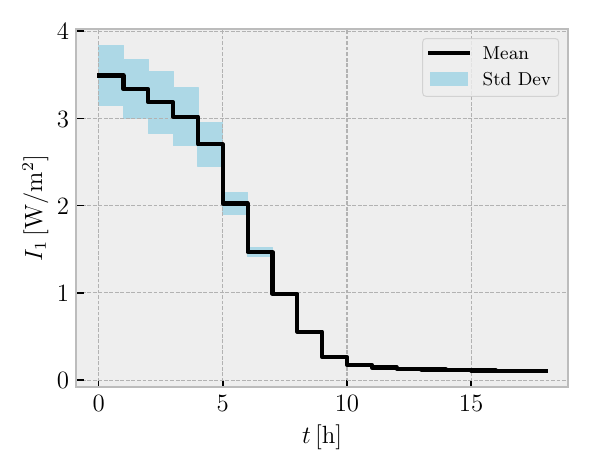}
      \subcaption[]{Input 1}
    \end{subfigure}
    %%%%%% INPUT 2 %%%%%% 
    \begin{subfigure}{0.24\textwidth}
      \captionsetup{justification=centering}
      \includegraphics[scale=0.4]{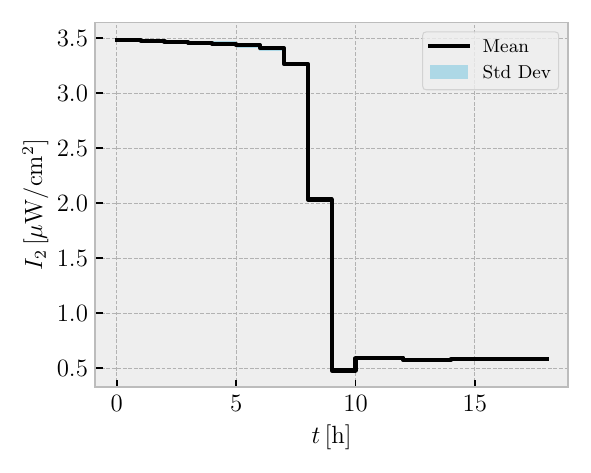}
      \subcaption[]{Input 2}
    \end{subfigure}
    \begin{subfigure}{0.24\textwidth}
      \captionsetup{justification=centering}
      \includegraphics[scale=0.4]{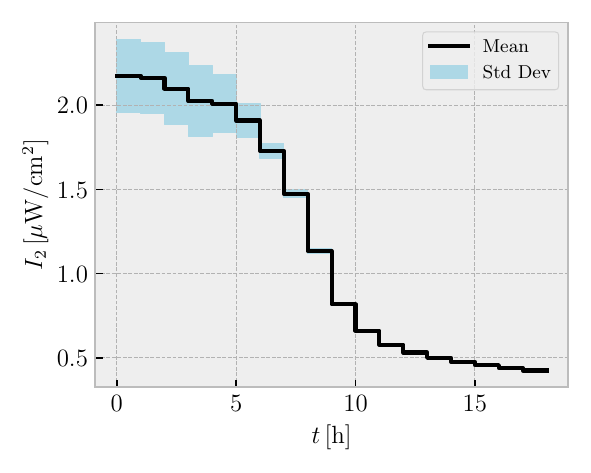}
      \subcaption[]{Input 2}
    \end{subfigure}
    \vline
    \begin{subfigure}{0.24\textwidth}
      \captionsetup{justification=centering}
      \includegraphics[scale=0.4]{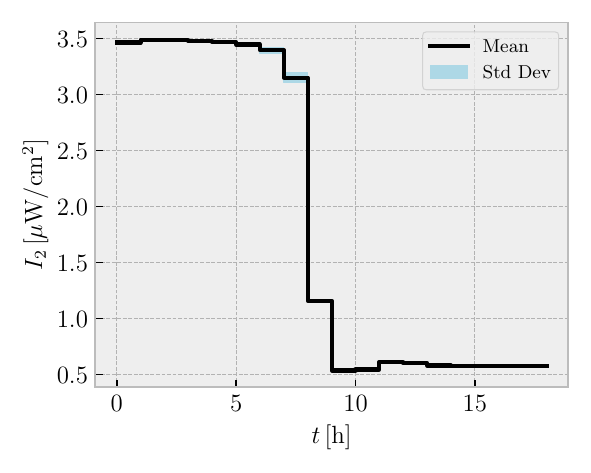}
      \subcaption[]{Input 2}
    \end{subfigure}
    \begin{subfigure}{0.24\textwidth}
      \captionsetup{justification=centering}
      \includegraphics[scale=0.4]{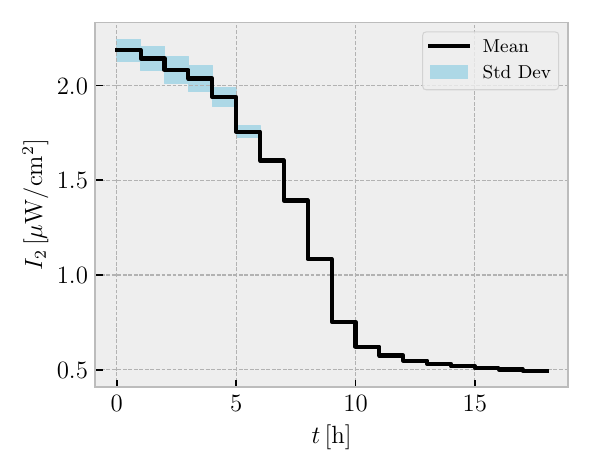}
      \subcaption[]{Input 2}
    \end{subfigure}
    \caption{Results for control case 1 (multi-setpoint tracking \textit{without} uncertainty) for two selected setpoint combinations ($b_1^*, b_2^*$): $(3,4)$ and $(3.5,3.5)$. The \textit{normalized} return function $J^*$, scaled to the range $[0,1]$ based on the maximum value achieved, is plotted over all epochs until early stopping occurred or the maximum number of epochs was reached. Dynamic plots for biomass concentrations, growth rates, and applied inputs correspond to the epoch with the maximum mean return function value (red mark in the plot of the return function). The dotted red lines in the biomass plots represent the target setpoints, while the dotted red line in the plots of the growth rate represents the bioreactor's dilution rate. The blue shaded area indicates the standard deviation.}
    \label{fig:control_case_1_plots_part2}
\end{figure*}
\FloatBarrier
\clearpage 

Considering four setpoint combinations ($b_1^*, b_2^*$), namely $(1,6)$, $(2,5)$, $(3,4)$, and $(3.5,3.5)$, Figs. \ref{fig:control_case_1_plots_part1} and \ref{fig:control_case_1_plots_part2} present the best-performing scenarios for control case 1. We compare the results against the benchmark quadratic-cost-based return function. The dynamic plots correspond to the epoch with the highest mean return function value in the respective scenario. As shown, the RL agent, using our proposed saturation-based return function, successfully tracks all setpoints by dynamically modulating the growth rates.

To better \textit{interpret} the actions of the RL agent using our proposed approach, we can see that once the target biomass concentrations are reached, the growth rates rapidly stabilize at or close to the bioreactor’s dilution rate, preventing further biomass accumulation. In other words, the RL agent focuses on rapidly reaching the biomass population targets during the \textit{transient} phase of the process, then shifts its focus to maintaining the biomass at the setpoints during \textit{steady-state} operation. In addition, with the saturation-based return function, the return values increase smoothly over epochs without aggressive jumps or oscillations, as expected given the smoother gradients. Despite the controlled system being deterministic in control case 1, the stochastic policy facilitates \textit{natural} exploration before converging to a more deterministic behavior. 

In contrast, the benchmark quadratic-cost-based return function fails to achieve proper setpoint tracking, particularly for the setpoints $(2,5)$, $(3,4)$, and $(3.5,3.5)$. In the latter case, the systems exhibit an initial overshoot followed by an undershoot without achieving actual convergence. Similarly, the growth rates fail to stabilize near the bioreactor’s dilution rate upon reaching the target population levels, which explains the poor tracking performance. Comparatively, for setpoint $(1,6)$, the quadratic-cost-based return function does guide the biomass concentrations closer to the targets, but our proposed saturation-based return function still achieves better tracking performance and does so slightly earlier in time. {The superiority of our saturation-based return function is also demonstrated in Table \ref{tab:metrics_total_naae}, where the total NAAE for the proposed return function is always smaller than that of the quadratic-cost-based counterpart.}

\begin{table}[ht]
\centering
\caption{{Total normalized average absolute error (NAAE) for the epoch that achieves the highest mean return in each scenario in cases 1–4. Results are shown for both the saturation-based return (SBR) and the quadratic-cost-based return (QBR). To capture variability, we modified Eq. \eqref{eq:naae_i_av} so that NAAE is first computed per episode and then averaged across all episodes within that epoch.}}
\label{tab:metrics_total_naae}
\begin{tabular}{@{}ccrrrr@{}}
\toprule
\multirow{2}{*}{\textbf{Case}} & \multirow{2}{*}{\textbf{Scenario}} &
\multicolumn{2}{c}{\textbf{SBR}} & \multicolumn{2}{c}{\textbf{QBR}} \\
\cmidrule(lr){3-4}\cmidrule(l){5-6}
 & & \multicolumn{1}{c}{\textbf{mean}} & \multicolumn{1}{c}{\textbf{std}} &
   \multicolumn{1}{c}{\textbf{mean}} & \multicolumn{1}{c}{\textbf{std}} \\ \midrule
\multirow{4}{*}{1} & Setpoint: $b_1^* = 1, b_2^* = 6$, no uncertainty & 0.394 & 0.001 & 0.450 & 0.002 \\
                   & Setpoint: $b_1^* = 2, b_2^* = 5$, no uncertainty & 0.393 & 0.000 & 0.444 & 0.001 \\
                   & Setpoint: $b_1^* = 3, b_2^* = 4$, no uncertainty & 0.391 & 0.000 & 0.468 & 0.003 \\
                   & Setpoint: $b_1^* = 3.5, b_2^* = 3.5$, no uncertainty & 0.394 & 0.000 & 0.446 & 0.000 \\ \midrule
\multirow{2}{*}{2} & Trajectory $\phi = 0.5$, no uncertainty & 0.007 & 0.000 & 1.467 & 0.000
 \\
                   & Trajectory $\phi = 0.7$, no uncertainty & 0.009 & 0.000 & 1.465 & 0.002 \\ \midrule
                   
\multirow{1}{*}{3} & {Setpoint: $b_1^* = 3, b_2^* = 4$, uncertainty: 7 \%} & 0.430 & 0.024 & 0.506 & 0.022 \\ \midrule
\multirow{1}{*}{4} & Trajectory $\phi = 0.7$, uncertainty: 7 \% & 0.032 & 0.012 & 1.465 & 0.022 \\  \bottomrule
\end{tabular} 
\end{table}

Moreover, the return function in all the benchmark scenarios oscillates more aggressively and/or shows stagnant learning over large segments of training epochs. This contrasts with the saturation-based return function, which leads to smoother and faster learning dynamics. Overall, this demonstrates the added value of our proposed RL approach for multi-setpoint RL schemes. It offers both improved control compliance, as well as more stable and efficient learning.

\subsubsection{Case 2: multi-trajectory tracking without uncertainty}
\begin{figure*}[h!]
    \centering
    %%%%%% REWARDS %%%%%% 
    \makebox[0.48\textwidth][c]{\textbf{Trajectory $\phi = 0.5$}, \textbf{no uncertainty}}
    \makebox[0.48\textwidth][c]{\textbf{Trajectory $\phi = 0.7$}, \textbf{no uncertainty}}\\
    \makebox[0.24\textwidth][c]{\textbf{Exp.}: 1\_sr\_1\_tr\_$\beta$\_27}
    \makebox[0.24\textwidth][c]{\textbf{Exp.}: qc}
    \makebox[0.24\textwidth][c]{\textbf{Exp.}: 1\_sr\_1\_tr\_$\beta$\_27}
    \makebox[0.24\textwidth][c]{\textbf{Exp.}: qc}\\
    \begin{subfigure}{0.24\textwidth}
      \captionsetup{justification=centering}
      \includegraphics[scale=0.4]{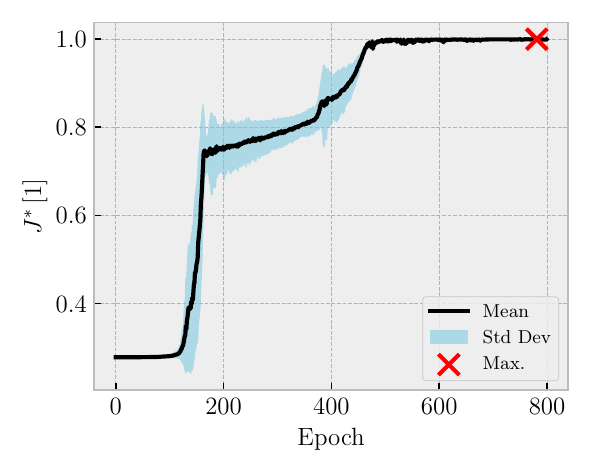}
      \subcaption[]{Reward}
    \end{subfigure}
    \begin{subfigure}{0.24\textwidth}
      \captionsetup{justification=centering}
      \includegraphics[scale=0.4]{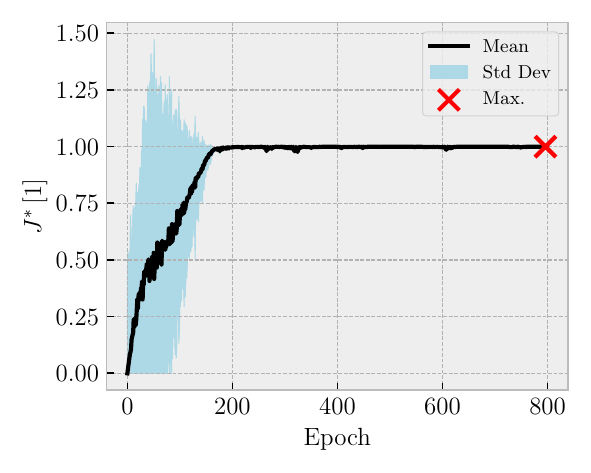}
      \subcaption[]{Reward}
    \end{subfigure}
    \vline
    \begin{subfigure}{0.24\textwidth}
      \captionsetup{justification=centering}
      \includegraphics[scale=0.4]{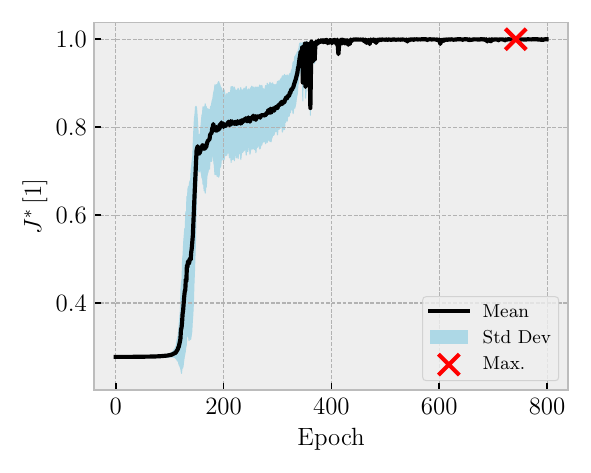}
      \subcaption[]{Reward}
    \end{subfigure}
    \begin{subfigure}{0.24\textwidth}
      \captionsetup{justification=centering}
      \includegraphics[scale=0.4]{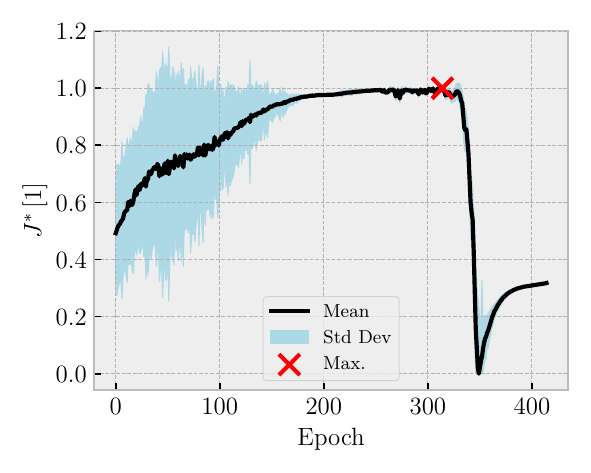}
      \subcaption[]{Reward}
    \end{subfigure}
    %%%%%% BIOMASS %%%%%% 
    \begin{subfigure}{0.24\textwidth}
      \captionsetup{justification=centering}
      \includegraphics[scale=0.4]{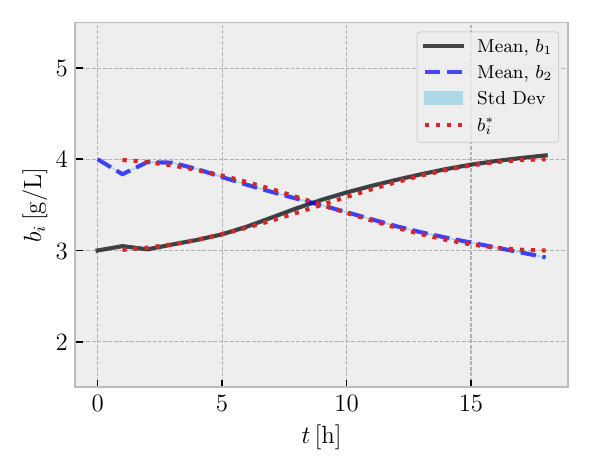}
      \subcaption[]{Biomass}
    \end{subfigure}
    \begin{subfigure}{0.24\textwidth}
      \captionsetup{justification=centering}
      \includegraphics[scale=0.4]{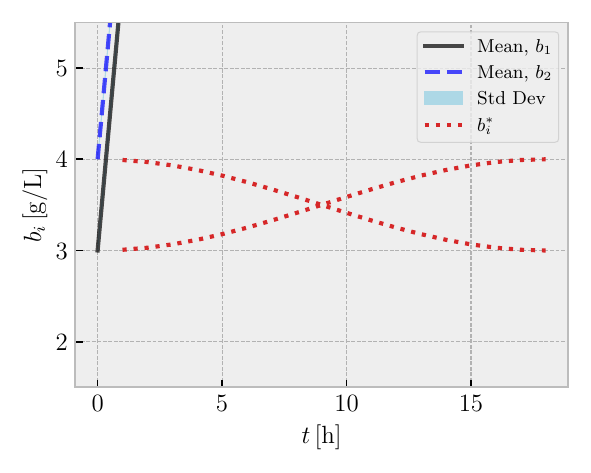}
      \subcaption[]{Biomass}
    \end{subfigure}
    \vline
    \begin{subfigure}{0.24\textwidth}
      \captionsetup{justification=centering}
      \includegraphics[scale=0.4]{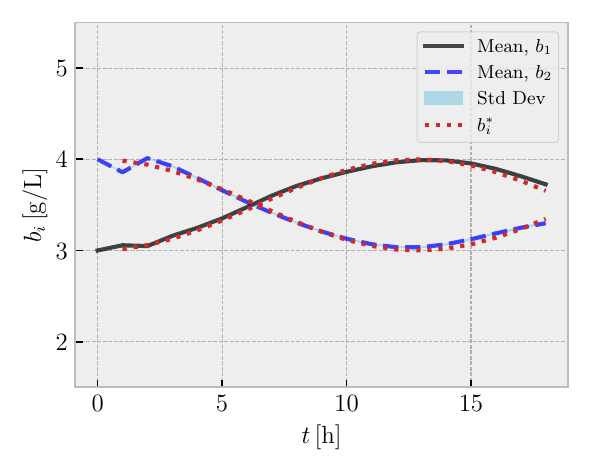}
      \subcaption[]{Biomass}
    \end{subfigure}
    \begin{subfigure}{0.24\textwidth}
      \captionsetup{justification=centering}
      \includegraphics[scale=0.4]{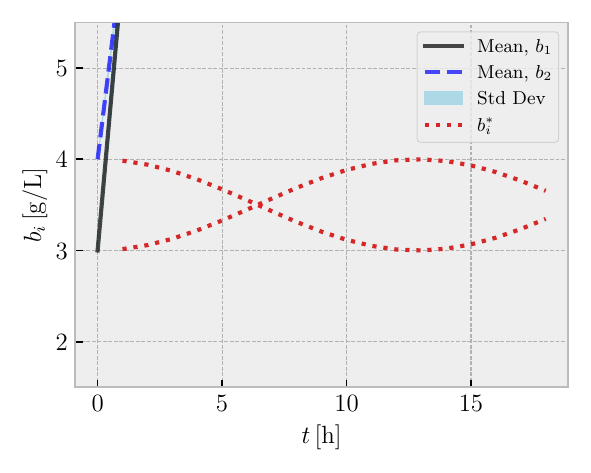}
      \subcaption[]{Biomass}
    \end{subfigure}
    %%%%%% GROWTH RATE %%%%%% 
    \begin{subfigure}{0.24\textwidth}
      \captionsetup{justification=centering}
      \includegraphics[scale=0.4]{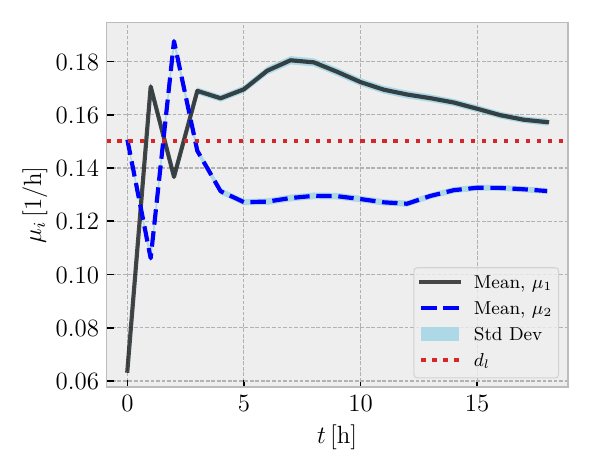}
      \subcaption[]{Growth rates}
    \end{subfigure}
    \begin{subfigure}{0.24\textwidth}
      \captionsetup{justification=centering}
      \includegraphics[scale=0.4]{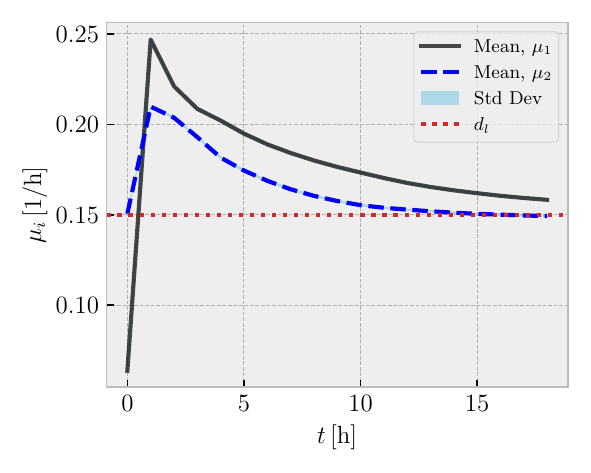}
      \subcaption[]{Growth rates}
    \end{subfigure}
    \vline
    \begin{subfigure}{0.24\textwidth}
      \captionsetup{justification=centering}
      \includegraphics[scale=0.4]{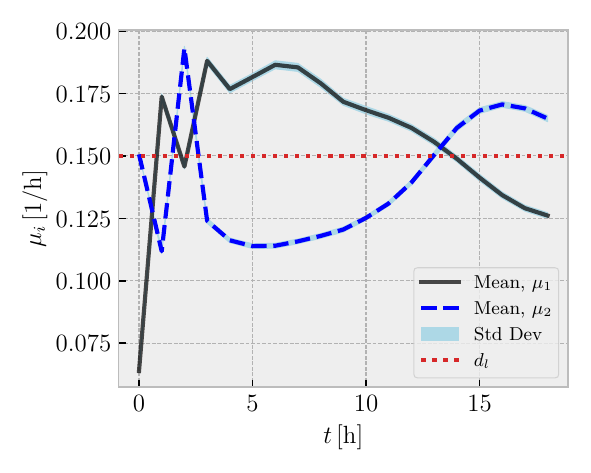}
      \subcaption[]{Growth rates}
    \end{subfigure}
    \begin{subfigure}{0.24\textwidth}
      \captionsetup{justification=centering}
      \includegraphics[scale=0.4]{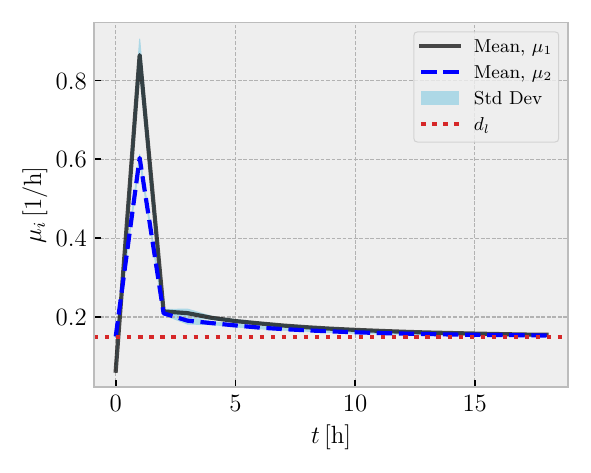}
      \subcaption[]{Growth rates}
    \end{subfigure}
    %%%%%% INPUT 1 %%%%%% 
    \begin{subfigure}{0.24\textwidth}
      \captionsetup{justification=centering}
      \includegraphics[scale=0.4]{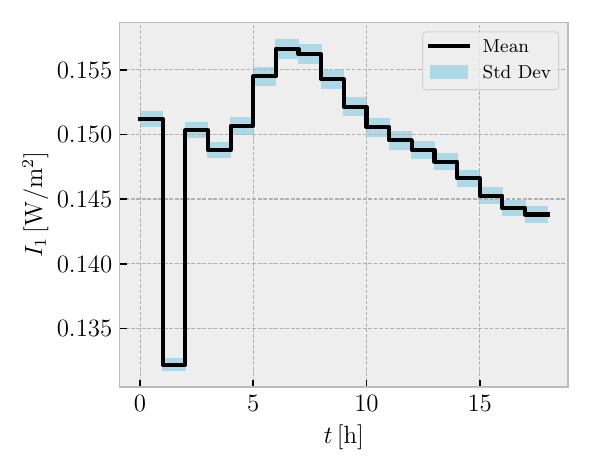}
      \subcaption[]{Input 1}
    \end{subfigure}
    \begin{subfigure}{0.24\textwidth}
      \captionsetup{justification=centering}
      \includegraphics[scale=0.4]{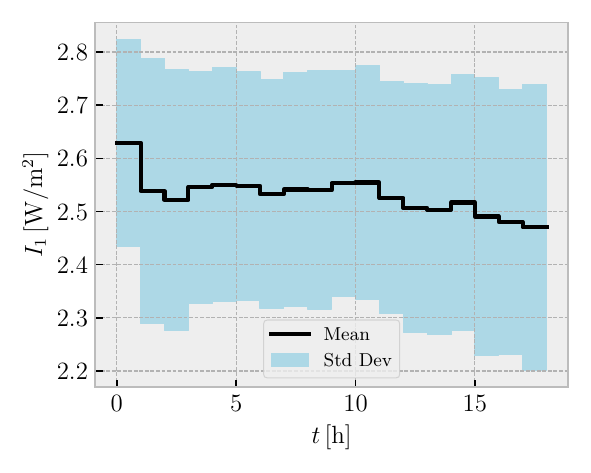}
      \subcaption[]{Input 1}
    \end{subfigure}
    \vline
    \begin{subfigure}{0.24\textwidth}
      \captionsetup{justification=centering}
      \includegraphics[scale=0.4]{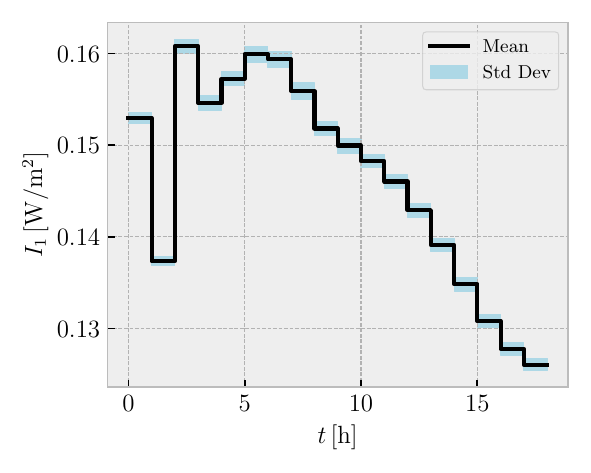}
      \subcaption[]{Input 1}
    \end{subfigure}
    \begin{subfigure}{0.24\textwidth}
      \captionsetup{justification=centering}
      \includegraphics[scale=0.4]{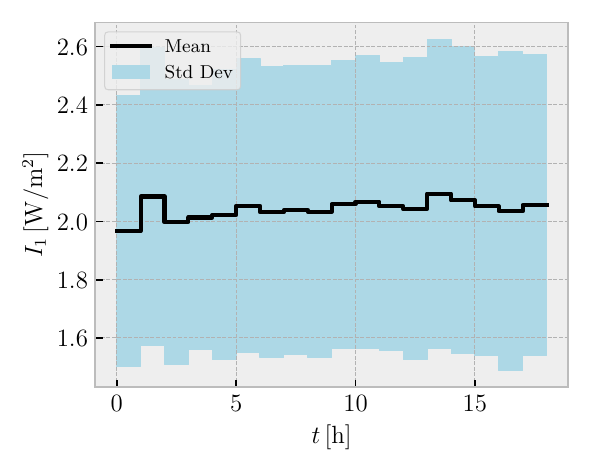}
      \subcaption[]{Input 1}
    \end{subfigure}
    %%%%%% INPUT 2 %%%%%% 
    \begin{subfigure}{0.24\textwidth}
      \captionsetup{justification=centering}
      \includegraphics[scale=0.4]{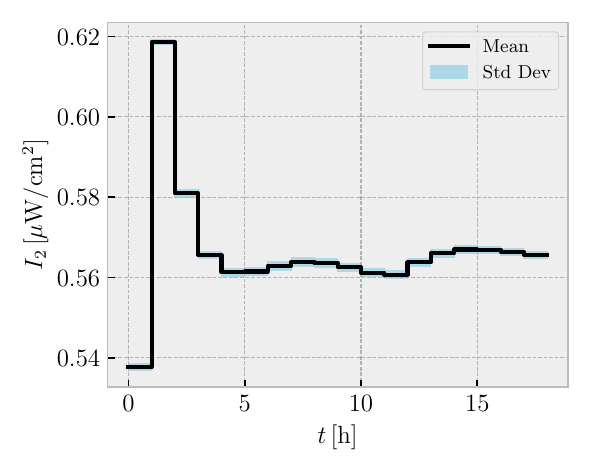}
      \subcaption[]{Input 2}
    \end{subfigure}
    \begin{subfigure}{0.24\textwidth}
      \captionsetup{justification=centering}
      \includegraphics[scale=0.4]{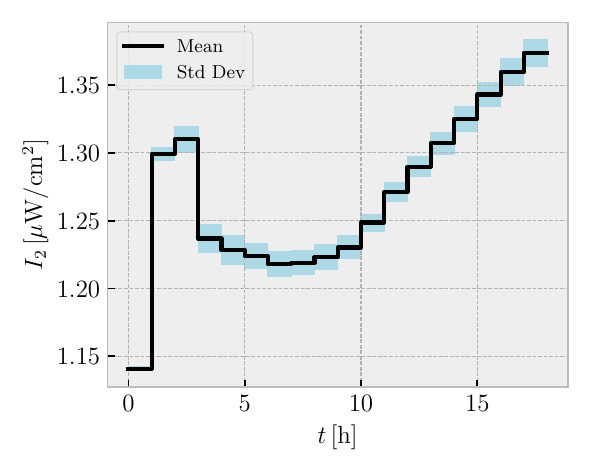}
      \subcaption[]{Input 2}
    \end{subfigure}
    \vline
    \begin{subfigure}{0.24\textwidth}
      \captionsetup{justification=centering}
      \includegraphics[scale=0.4]{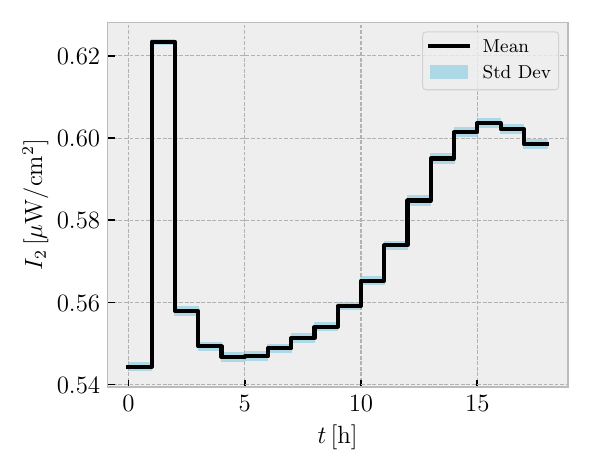}
      \subcaption[]{Input 2}
    \end{subfigure}
    \begin{subfigure}{0.24\textwidth}
      \captionsetup{justification=centering}
      \includegraphics[scale=0.4]{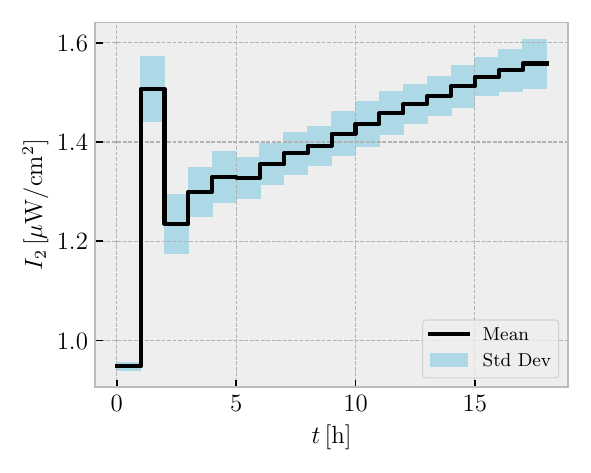}
      \subcaption[]{Input 2}
    \end{subfigure}
    \caption{Results for control case 2 (multi-trajectory tracking \textit{without} uncertainty) for two selected smooth sinusoidal trajectories ($b_1^*, b_2^*$). The \textit{normalized} return function $J^*$, scaled to the range $[0,1]$ based on the maximum value achieved, is plotted over all epochs until early stopping occurred or the maximum number of epochs was reached. Dynamic plots for biomass concentrations, growth rates, and applied inputs correspond to the epoch with the maximum mean return function value (red mark in the plot of the return function). The dotted red lines in the biomass plots represent the target trajectories, while the dotted red line in the plots of the growth rate represents the bioreactor's dilution rate. The blue shaded area indicates the standard deviation.}
    \label{fig:control_case_2_plots}
\end{figure*}

Compared to the multi-setpoint tracking task in control case 1, control case 2 is inherently more complex. As shown in Fig. \ref{fig:control_case_2_plots}, we tested two multi-trajectory combinations, where the reference setpoints $(b_1^*, b_2^*)$ are \textit{dynamic} rather than constant. The reference signals were designed as smooth sinusoidal trajectories oscillating between 3 and 4. The two experiments differ in the frequency of oscillation \(\phi\) (i.e., the number of cycles within the total time horizon), namely $\phi = 0.5$ and $\phi = 0.7$. The saturation-based return function was shaped using an equal-stage-terminal reward scheme with $\beta_{\epsilon_i} = 27$, i.e., the best configuration shown in Fig. \ref{fig:bar_plots_rank}-C. We considered $N_m = 800$ epochs, 300 more than in control case 1, due to the added complexity of the dynamic multi-trajectory tracking task.

\FloatBarrier
\clearpage 
\FloatBarrier
\clearpage

The results indicate that our proposed saturation-based return function enables efficient multi-trajectory tracking, whereas the benchmark quadratic-cost-based function fails to converge to an acceptable solution, deviating significantly from the desired trajectories. {This is also evident in Table \ref{tab:metrics_total_naae}, where the total NAAE achieved by the saturation-based return function is significantly lower than that obtained with the quadratic-cost-based counterpart.} Unlike in control case 1, where the quadratic-cost function \textit{at least} approximated the reference setpoint values, it completely fails in this more complex task. Notably, achieving convergence with our saturation-based return function required more training epochs than in control case 1, justifying the increased maximum number of epochs. The good performance of our proposed return function for multi-trajectory tracking can be interpreted as the agent's ability to modulate growth rates effectively, raising them above the bioreactor’s dilution rate when biomass is expected to increase, and lowering them when biomass is expected to decrease. In contrast, the quadratic-cost-based function produced excessive growth rates from the start, well above the dilution rate, causing significant drift of the biomass population levels from the dynamic reference trajectories.

Overall, the results from control cases 1 and 2 demonstrate that our proposed saturation-based return function is effective for both multi-setpoint and multi-trajectory tracking tasks, while the benchmark quadratic-cost-based function shows limited success in multi-setpoint tracking and fails entirely in multi-trajectory tracking.

\subsubsection{Case 3: robust multi-setpoint tracking under uncertainty}
\label{subsubsub:case_3}
We incorporated system uncertainty into the multi-setpoint tracking task to evaluate the robustness of our method. Specifically, we introduced a 7 \% error in all initial conditions and in two key parameters that directly influence the input-dependent production rates of the amino acids regulating auxotrophic growth in the consortium, namely $q_{a_{\max_1}}$ and $q_{a_{\max_2}}$. These uncertain parameters were sampled from Gaussian distributions during Monte Carlo simulations, using the nominal values in Table \ref{tab:parameters} as means and a 7 \% standard deviation. This level of uncertainty introduces significant variability into the system, making the learning task more challenging and providing a strong test case for evaluating robustness. To ensure \textit{controlled} randomization, we truncated the distribution at three standard deviations, effectively covering $\sim 99.7 \, \%$ of the cumulative probability. As a proof of concept, we considered the setpoint combination ($b_1^*=3, b_2^*=4$) from control case 1, now under the outlined uncertain conditions.

\begin{figure*}[h!]
    \centering
    %%%%%% REWARDS %%%%%% 
    \makebox[0.48\textwidth][c]{{\textbf{Setpoint:} $b_1^* = 3, b_2^* = 4$, \textbf{uncertainty}: 7 \%}}\\
    \makebox[0.24\textwidth][c]{\textbf{Exp.}: 1\_sr\_1\_tr\_$\beta$\_27}
    \makebox[0.24\textwidth][c]{\textbf{Exp.}: qc}
    
    \begin{subfigure}{0.24\textwidth}
      \captionsetup{justification=centering}
      \includegraphics[scale=0.4]{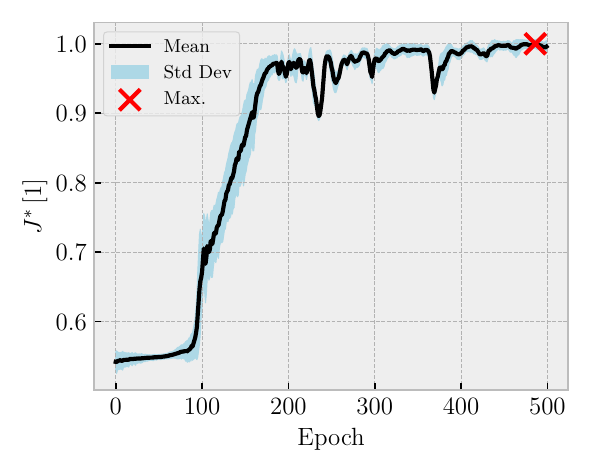}
      \subcaption[]{Reward}
    \end{subfigure}
    \begin{subfigure}{0.24\textwidth}
      \captionsetup{justification=centering}
      \includegraphics[scale=0.4]{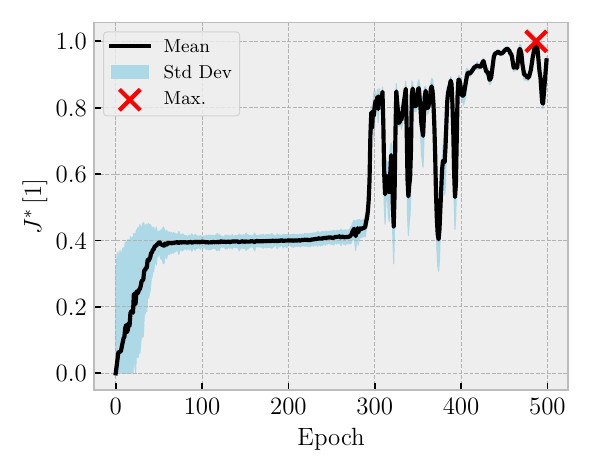}
      \subcaption[]{Reward}
    \end{subfigure}
    
    %%%%%% BIOMASS %%%%%% 
    \begin{subfigure}{0.24\textwidth}
      \captionsetup{justification=centering}
      \includegraphics[scale=0.4]{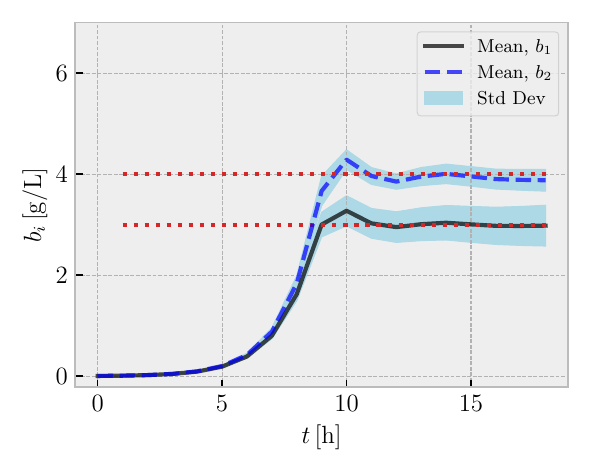}
      \subcaption[]{Biomass}
    \end{subfigure}
    \begin{subfigure}{0.24\textwidth}
      \captionsetup{justification=centering}
      \includegraphics[scale=0.4]{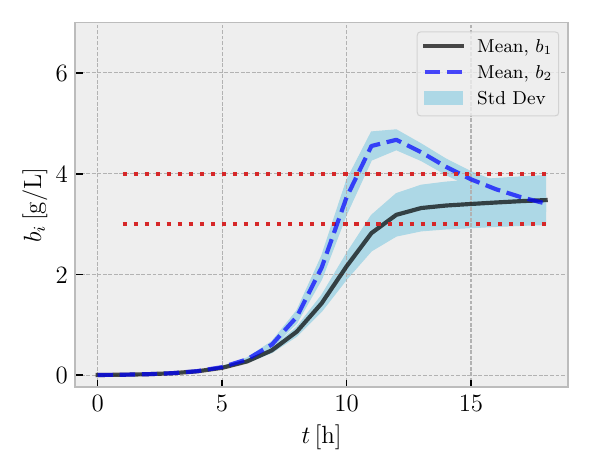}
      \subcaption[]{Biomass}
    \end{subfigure}
    
    %%%%%% GROWTH RATE %%%%%% 
    \begin{subfigure}{0.24\textwidth}
      \captionsetup{justification=centering}
      \includegraphics[scale=0.4]{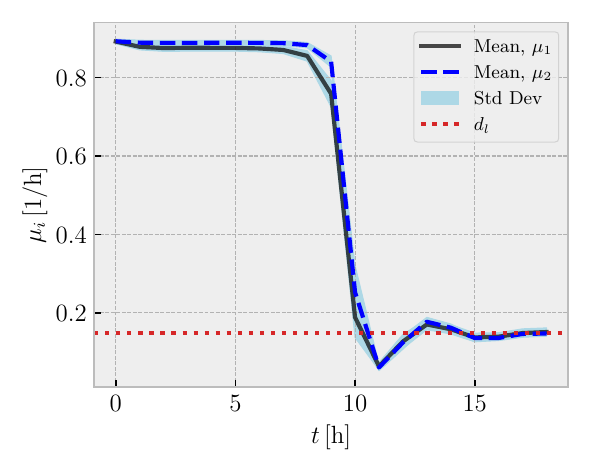}
      \subcaption[]{Growth rates}
    \end{subfigure}
    \begin{subfigure}{0.24\textwidth}
      \captionsetup{justification=centering}
      \includegraphics[scale=0.4]{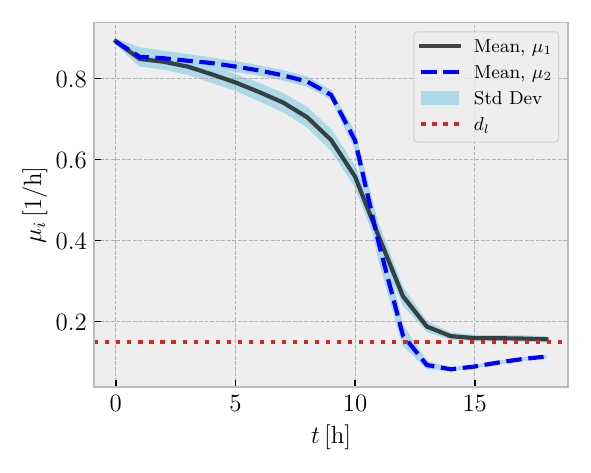}
      \subcaption[]{Growth rates}
    \end{subfigure}
    
    %%%%%% INPUT 1 %%%%%% 
    \begin{subfigure}{0.24\textwidth}
      \captionsetup{justification=centering}
      \includegraphics[scale=0.4]{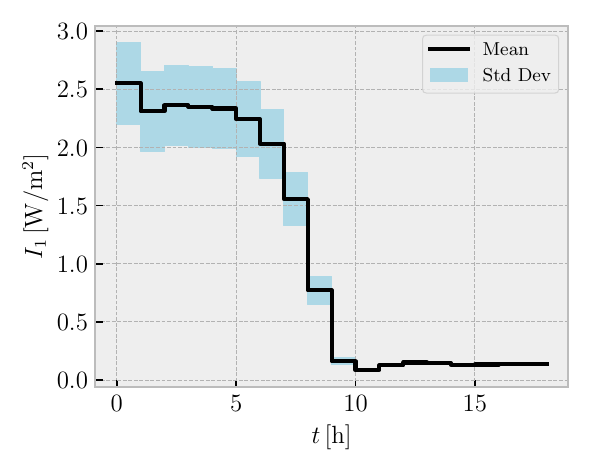}
      \subcaption[]{Input 1}
    \end{subfigure}
    \begin{subfigure}{0.24\textwidth}
      \captionsetup{justification=centering}
      \includegraphics[scale=0.4]{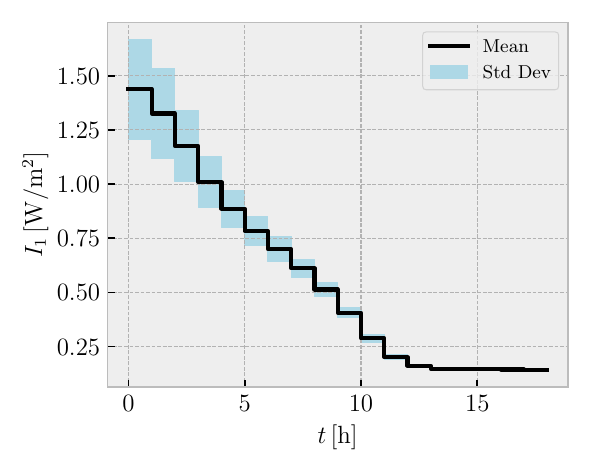}
      \subcaption[]{Input 1}
    \end{subfigure}
    
    %%%%%% INPUT 2 %%%%%% 
    \begin{subfigure}{0.24\textwidth}
      \captionsetup{justification=centering}
      \includegraphics[scale=0.4]{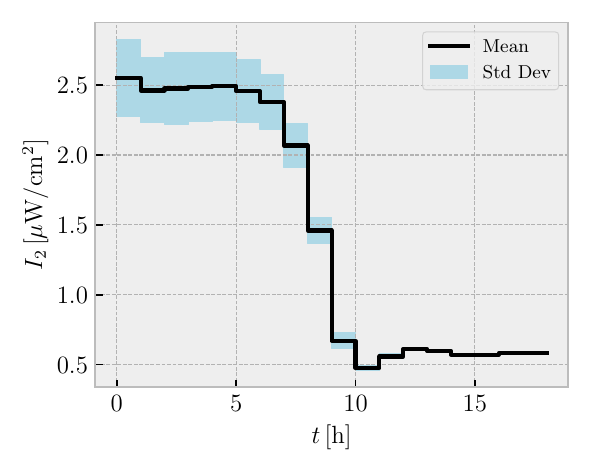}
      \subcaption[]{Input 2}
    \end{subfigure}
    \begin{subfigure}{0.24\textwidth}
      \captionsetup{justification=centering}
      \includegraphics[scale=0.4]{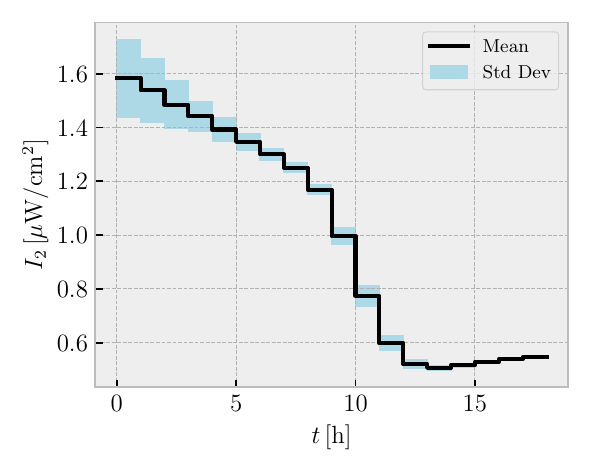}
      \subcaption[]{Input 2}
    \end{subfigure}
    \caption{Results for control case 3 (robust multi-setpoint tracking \textit{under} uncertainty) for the setpoint combination ($b_1^*=3, b_2^*=4$). The \textit{normalized} return function $J^*$, scaled to the range $[0,1]$ based on the maximum value achieved, is plotted over all epochs until early stopping occurred or the maximum number of epochs was reached. Dynamic plots for biomass concentrations, growth rates, and applied inputs correspond to the epoch with the maximum mean return function value (red mark in the plot of the return function). The dotted red lines in the biomass plots represent the target setpoints, while the dotted red line in the plots of the growth rate represents the bioreactor's dilution rate. The blue shaded area indicates the standard deviation.}
    \label{fig:control_case_3_plots}
\end{figure*}

The results in Fig. \ref{fig:control_case_3_plots} demonstrate that our proposed saturation-based return function enables efficient multi-setpoint tracking \textit{on average} under uncertainty, i.e., it exhibits robustness, as the \textit{mean} trajectory closely follows the defined setpoints. Naturally, while the mean trajectories exhibit similar trends to those observed in the case without system uncertainty (cf. Fig. \ref{fig:control_case_1_plots_part2}), a higher standard deviation is evident due to the embedded uncertainty in the initial conditions and selected parameters. In contrast, the quadratic-cost-based function fails to accurately track the multiple setpoints under uncertainty, consistent with its performance in the previous evaluation without uncertainty. {The poorer performance of the quadratic-cost-based return function is also shown in Table \ref{tab:metrics_total_naae}, where its total NAAE is higher than that obtained with the saturation-based one.} Another notable aspect is the behavior of the return function over epochs. With our saturation-based function, the learning process remains relatively stable, showing only slight oscillations as the improvement rate slows down. In contrast, the quadratic-cost-based function exhibits less stable learning, plateauing for a significant period before experiencing abrupt oscillations after approximately 300 epochs.

\subsubsection{Case 4: robust multi-trajectory tracking under uncertainty}
\label{subsubsub:case_4}
\begin{figure}[h!]
    \centering
    %%%%%% REWARDS %%%%%% 
    \makebox[0.48\textwidth][c]{{\textbf{Trajectory $\phi = 0.7$}, \textbf{uncertainty}: 7 \%}}\\
    \makebox[0.24\textwidth][c]{\textbf{Exp.}: 1\_sr\_1\_tr\_$\beta$\_27}
    \makebox[0.24\textwidth][c]{\textbf{Exp.}: qc}
    
    \begin{subfigure}{0.24\textwidth}
      \captionsetup{justification=centering}
      \includegraphics[scale=0.4]{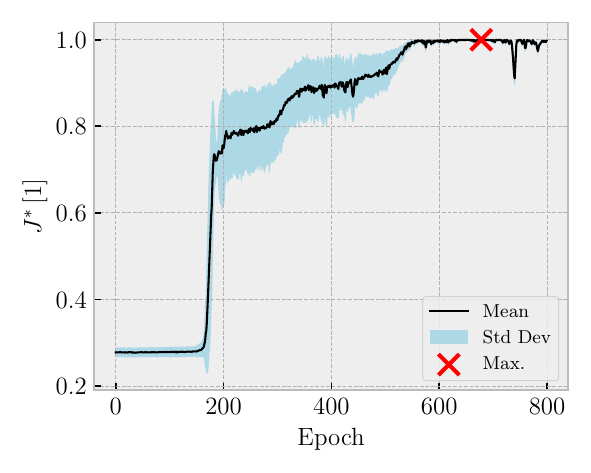}
      \subcaption[]{Reward}
    \end{subfigure}
    \begin{subfigure}{0.24\textwidth}
      \captionsetup{justification=centering}
      \includegraphics[scale=0.4]{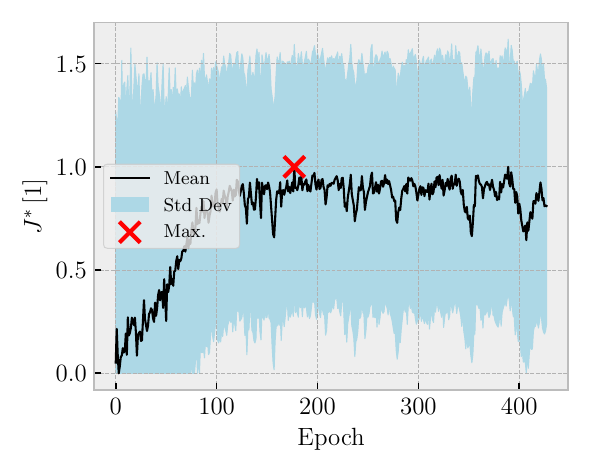}
      \subcaption[]{Reward}
    \end{subfigure}

    %%%%%% BIOMASS %%%%%% 
    \begin{subfigure}{0.24\textwidth}
      \captionsetup{justification=centering}
      \includegraphics[scale=0.4]{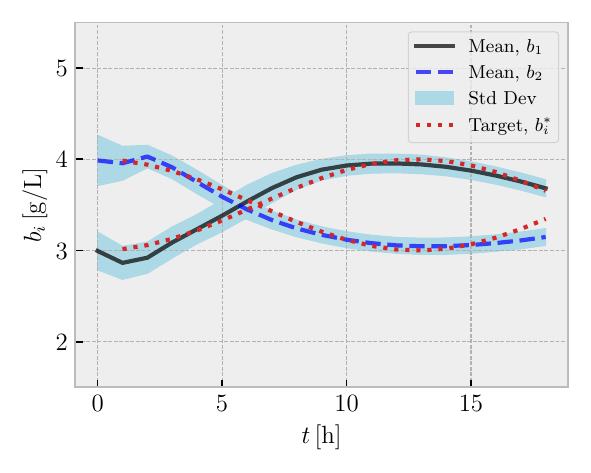}
      \subcaption[]{Biomass}
    \end{subfigure}
    \begin{subfigure}{0.24\textwidth}
      \captionsetup{justification=centering}
      \includegraphics[scale=0.4]{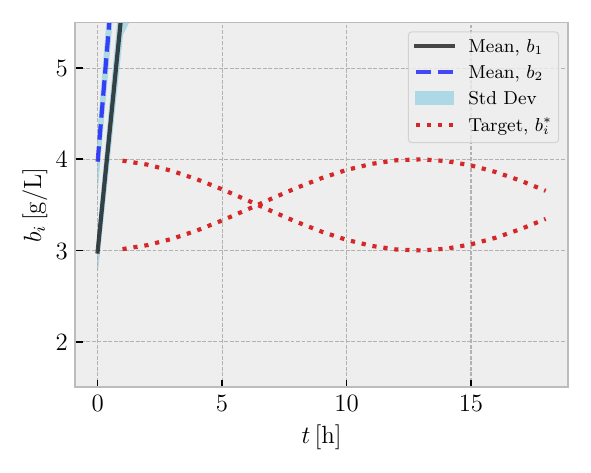}
      \subcaption[]{Biomass}
    \end{subfigure}

    %%%%%% GROWTH RATE %%%%%% 
    \begin{subfigure}{0.24\textwidth}
      \captionsetup{justification=centering}
      \includegraphics[scale=0.4]{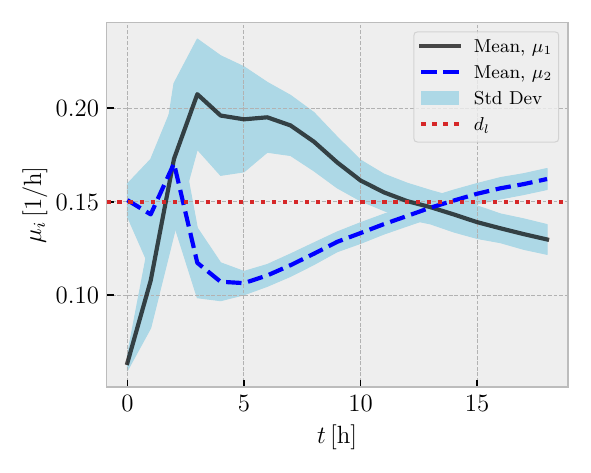}
      \subcaption[]{Growth rates}
    \end{subfigure}
    \begin{subfigure}{0.24\textwidth}
      \captionsetup{justification=centering}
      \includegraphics[scale=0.4]{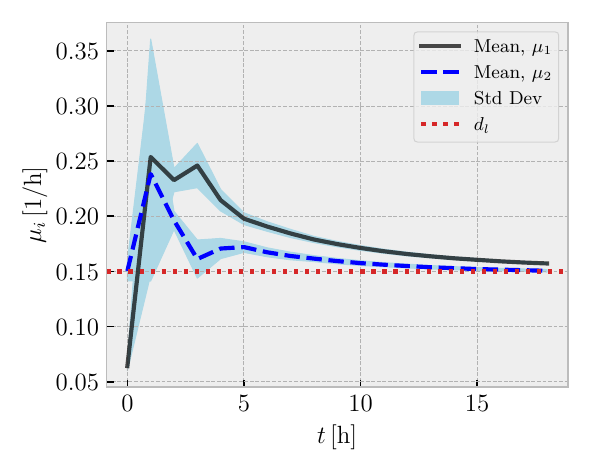}
      \subcaption[]{Growth rates}
    \end{subfigure}

    %%%%%% INPUT 1 %%%%%% 
    \begin{subfigure}{0.24\textwidth}
      \captionsetup{justification=centering}
      \includegraphics[scale=0.4]{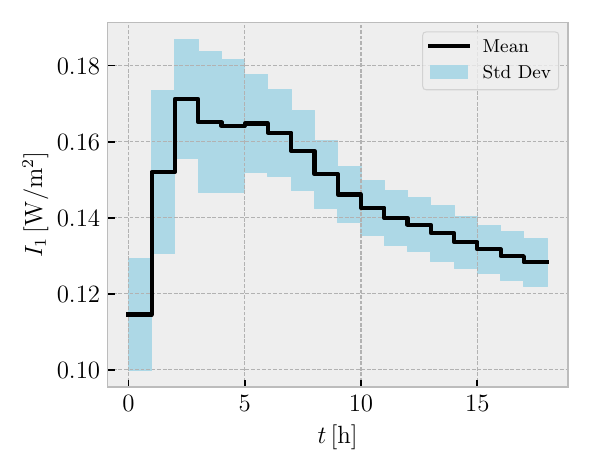}
      \subcaption[]{Input 1}
    \end{subfigure}
    \begin{subfigure}{0.24\textwidth}
      \captionsetup{justification=centering}
      \includegraphics[scale=0.4]{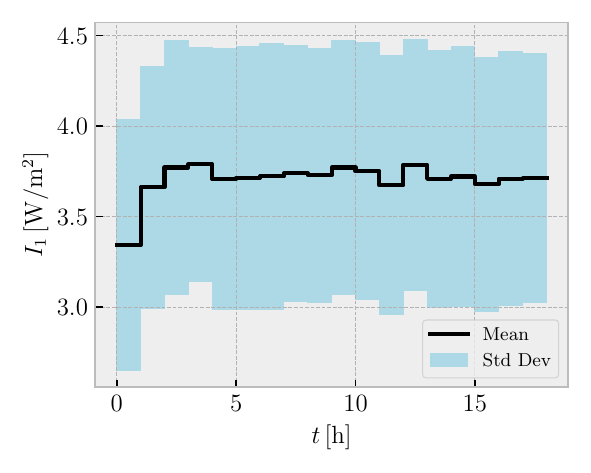}
      \subcaption[]{Input 1}
    \end{subfigure}

    %%%%%% INPUT 2 %%%%%% 
    \begin{subfigure}{0.24\textwidth}
      \captionsetup{justification=centering}
      \includegraphics[scale=0.4]{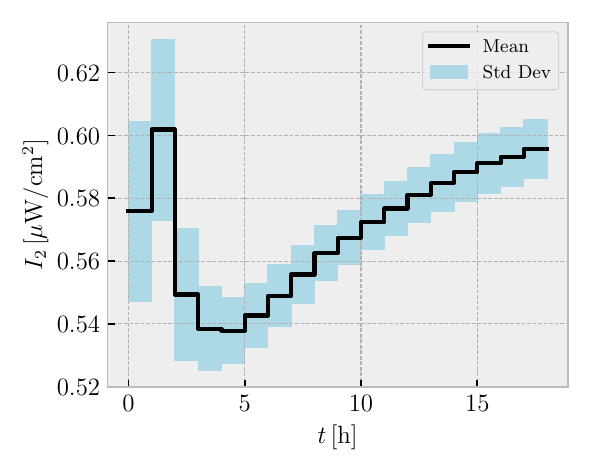}
      \subcaption[]{Input 2}
    \end{subfigure}
    \begin{subfigure}{0.24\textwidth}
      \captionsetup{justification=centering}
      \includegraphics[scale=0.4]{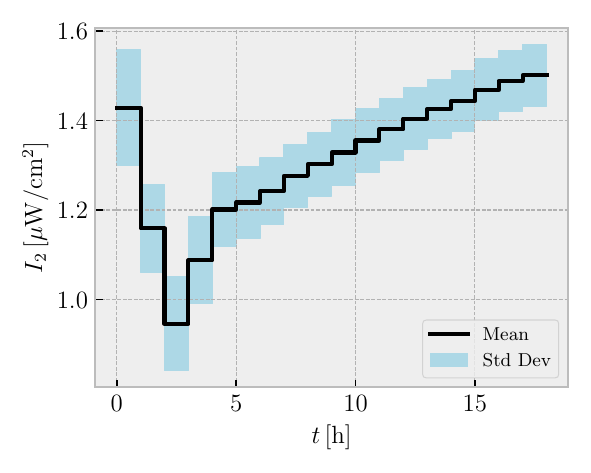}
      \subcaption[]{Input 2}
    \end{subfigure}
    \caption{Results for control case 4 (robust multi-trajectory tracking \textit{under} uncertainty) for a selected smooth sinusoidal trajectory ($b_1^*, b_2^*$). The \textit{normalized} return function $J^*$, scaled to the range $[0,1]$ based on the maximum value achieved, is plotted over all epochs until early stopping occurred or the maximum number of epochs was reached. Dynamic plots for biomass concentrations, growth rates, and applied inputs correspond to the epoch with the maximum mean return function value (red mark in the plot of the return function). The dotted red lines in the biomass plots represent the target trajectories, while the dotted red line in the plots of the growth rate represents the bioreactor's dilution rate. The blue shaded area indicates the standard deviation.}
    \label{fig:control_case_4_plots}
\end{figure}

We evaluated the performance of multi-trajectory tracking for the smooth sigmoidal trajectories with $\phi = 0.7$ tested in control case 2 (cf. Fig. \ref{fig:control_case_2_plots}), while applying the same uncertain conditions as in control case 3. As shown in Fig. \ref{fig:control_case_4_plots}, our proposed saturation-based return function successfully tracked the dynamic reference trajectories \textit{on average} despite uncertainty in the initial conditions and key parameters. The \textit{mean} biomass populations closely followed the reference trajectories, demonstrating robustness. As in control case 3, an increased standard deviation due to system uncertainty was observed. In contrast, the quadratic-cost-based return function failed to guide the agent toward a \textit{viable} policy, with trajectories deviating significantly from the reference, mirroring the poor performance observed in control case 2. {As with cases 1-3, the total NAAE in Table \ref{tab:metrics_total_naae} for the saturation-based return function is significantly lower that that of the quadratic-cost counterpart.}

\textit{Remark}. We also tested other uncertainty levels ranging from 1 \% to 7 \% error in Sections \ref{subsubsub:case_3} and \ref{subsubsub:case_4}, and the results were equally robust to those already presented. For conciseness, we only show the scenarios corresponding to the highest uncertainty level tested in this work.

\FloatBarrier
\clearpage
\FloatBarrier
\clearpage

\FloatBarrier
\clearpage

{The observed robustness against uncertainty in cases 3 and 4 is explained by the fact that the policy sees perturbed trajectories $\bm{\tau}$ during training. In other words, the RL agent experiences a wide range of possible dynamic behaviors over the Monte Carlo simulations, via domain randomization. Given that the RL agent maximizes the expected return $\mathbb{E}_{\bm{\tau}} \left[ J_s(\bm{\tau}) \right]$ (cf. Eq. \ref{eq:grad_exp_2}) during training, it adjusts its parameters toward favoring actions that perform well \textit{on average} across the entire uncertainty envelope. This expectation-driven learning enables robustness in the controller, as the control policy is able to \textit{anticipate} future uncertainties.}

Overall, these results confirm that our saturation-based return function is robust to system uncertainty in both multi-setpoint and multi-trajectory control problems, consistently demonstrating significant improvement compared to conventional quadratic-based return functions. Our framework's robustness is particularly advantageous for bioprocesses, where uncertainty is commonplace and can arise from variability in initial conditions, disturbances, or stochastic system dynamics. For real-world implementation of RL, such as in multi-setpoint and multi-trajectory tracking tasks, system uncertainty can be accounted for by incorporating domain randomization, as done here, or by enabling the policy to \textit{experience} uncertainty through sufficient exploration. In either case, our results show that the outlined RL method can generate uncertainty-aware policies with enhanced learning stability, control compliance, and robustness. 

{The goal of this study was to present a practical strategy for applying RL to multi-setpoint and multi-trajectory control problems, such as those encountered in microbial consortia, thereby expanding the bioprocess-control toolbox. Determining the superiority of RL over other possible control approaches is beyond the scope of this study. However, in practical deployments we recommend a cross-method analysis that may include, e.g., adaptive model-based control (when a reliable model is available), simpler PID schemes, and an RL approach like the one described here. A holistic evaluation, considering robustness, performance, adaptability, implementation effort, and available resources, will ultimately determine the most appropriate control strategy for a given process.}

{Finally, all scenarios in cases 1-4 were trained with the same policy-gradient algorithm; only the return calculation differs. Because the time required to compute the return is negligible relative to the rest of the algorithm, the per-iteration computational cost is essentially identical for the saturation-based and quadratic-based approaches. What differs is the number of epochs required to obtain a high-quality policy. In our tested scenarios, the saturation-based return generally converges in fewer epochs than the quadratic-cost-based return, and in most instances the quadratic form does not even reach satisfactory performance. Although a systematic computational-time benchmark is beyond the scope of this study, these observations suggest that the proposed approach achieves a satisfactory policy with less overall computational effort (i.e., fewer epochs).}

\section{Conclusion}
\label{sec:conclusion}
In this work, we outlined the use of RL for efficient and robust multi-setpoint and multi-trajectory tracking in bioprocess control. We introduced a novel return function based on multiplicative reciprocal saturation functions that couples reward gains to the simultaneous satisfaction of multiple references, better guiding the RL agent’s learning process. Through a biotechnologically relevant case study involving a microbial consortium with cybergenetic growth control enabled by optogenetics, we demonstrated the benefits of our approach via computational experiments. Unlike conventional quadratic-cost-based return functions, which struggle to balance multiple objectives, our method ensures stable learning, faster convergence, and improved control performance. Additionally, by tuning the parameters of the saturation functions, one can adjust their smoothness or steepness, influencing gradient updates and shaping the overall learning process. 

We further demonstrated the ability of our framework to handle uncertainties such as variable initial conditions and intrinsically noisy kinetics, providing robustness, a desired feature in industrial bioprocesses. This strong probabilistic performance under uncertainty makes our RL control scheme well-suited for real-world bioprocess applications, paving the way for advanced and adaptive control strategies in biotechnology. Looking ahead, we are actively extending our framework to consider aspects such as policy generalization and observability constraints. {We also seek to experimentally validate our RL control approach in biotechnological processes of industrial relevance, leveraging the concept of division of labor for metabolic engineering in microbial consortia.} Finally, while this work focuses on bioprocess control, the proposed methods are generalizable to other applications in process and systems engineering, where similar multi-setpoint and multi-trajectory control challenges may arise.

% ---------- Nomenclature / List of Symbols ----------
\appendix
\section*{{Nomenclature}}
\renewcommand{\arraystretch}{1.15}
\begin{threeparttable}
\footnotesize
\begin{tabular}{>{\raggedright\arraybackslash}p{3cm} p{8cm}}
\toprule
%%%%%%
RL   & Reinforcement learning \\
PID  & Proportional–integral–derivative controller \\
MPC  &  Model predictive control \\
SBR  &  Saturation-based return \\
QBR  & Quadratic-cost-based return \\
NAAE & Normalised average absolute error \\
NAUC  & Normalized area under the (return) curve \\
%%%%%%
$\bm{x}_t$               & Full dynamic state vector at time $t$ \\
$\bm{u}_t$               & Action/input vector at time $t$ \\
$\bm{s}_t$               & Observation vector at time $t$ \\
$\bm{d}_t$               & Random disturbance vector at time $t$ \\
$R_{t+1}$               & System reward at time $t+1$ upon receiving action $\bm{u}_t$ \\
$\bm{\tau}$             & Joint trajectory of observed states, actions, and rewards \\
%%%%%%Bioprocess variables%%%%%%
$g$                    & Glucose concentration \\
$b_i$                  & Biomass concentration of strain $i$ \\
$a_i$                  & Intracellular concentration of auxotrophic amino acid in strain $i$ \\
$I_i$                  & Light intensity driving optogenetic module in strain $i$ \\
$d_l$                  & Chemostat dilution rate \\
$g_{\text{in}}$        & Feed glucose concentration \\
$d_{a_i}$              & Degradation rate of auxotrophic amino acid in strain $i$ \\
%%%%%%Kinetic rates%%%%%%
$\mu_i$                & Growth rate of strain $i$ \\
$q_{g_i}$              & Glucose uptake rate of strain $i$ \\
$q_{a_i}$              & Synthesis rate of auxotrophic amino acid in strain $i$ \\
%%%%%%Kinetic parameters%%%%%%
$\mu_{\max_i}$, $q_{a_{\max_i}}$          & Maximum rate constants of strain $i$ \\
$k_{g_i}$, $k_{a_i}$, $k_{I_i}$              & Saturation constants of strain $i$ \\
$Y_{g/b_i}$            & Yield of substrate on biomass of strain $i$ \\
$n_i$                  & Hill coefficient of strain $i$ \\
$f_c$                  & Conversion factor \\
%%%%%%RL / return-function notation%%%%%%
$ l_{s,q}$, $ e_{s,q}$  &  Quadratic-cost stage and terminal rewards, respectively \\
$ l_{s,c}$, $ e_{s,c}$  &  Saturation-cost stage and terminal rewards, respectively \\
$\bm{x}^*_t$, $\bm{x}^*_{N_s}$  & Reference state vectors \\
$\mathbf{Q},\mathbf{Q_T}$ & Weight matrices in QBR \\
$\epsilon_{i}$       & Squared tracking error of tracked state $i$ \\
$\beta_{\epsilon_i}$   & Error saturation constant of tracked state $i$ in SBR \\
$\alpha_{\max}$        & Maximum per-step reward in SBR \\
$w_t$, $w_{N_s}$       & Stage/terminal reward weights in SBR \\
$J_s(\bm{\tau})$       & Stochastic return over trajectory $\bm{\tau}$ \\
$\pi(\bm{u}_t\!\mid\!\bm{s}_t,\bm{\theta})$ & Stochastic policy \\
$\bm{\theta}$, $\Theta$          & Trainable policy parameters \\
$\bm{m}_t,\bm{\sigma}_t$& Mean and standard deviation of Gaussian policy at time $t$\\
$\Bar{J}_{s_m}$        & Mean of the returns in an epoch \\
$\sigma_{J_{s_m}}$     & Standard deviation of returns in an epoch \\
$\epsilon_{\text{mach}}$ & Machine epsilon \\
$\alpha$               & Learning rate \\
$N_{\text{MC}}$        & Monte Carlo trajectories (episodes) per epoch \\
$N_{m}$                & Number of training epochs \\
$\phi$                 & Oscillation frequency of reference trajectory \\
$J^*$        & Normalized return scaled to $[0,1]$  \\
%%%%%%Sets and indices%%%%%%
$\mathcal{X}_{\text{track}}$ & Set of states in $\bm{x}$ being tracked \\
$\mathbb{E}{[\cdot]}$ & General expectation operator \\
$\Bar{(\cdot)}$ & General mean operator \\
$|{(\cdot)}|$ & General absolute value operator \\
$m$                    & Epoch index \\
$k$                    & Episode index \\
\bottomrule
\end{tabular}
\end{threeparttable}
% ---------- End of Nomenclature ----------

\section*{Acknowledgment}
\label{sec:acknowledgment}
\noindent SER is part of the Advanced Engineering Biology Future Science Platform (AEB FSP). JLA was supported by US-NSF grant MCB-2300239.

\bibliographystyle{elsarticle-num} 
\bibliography{bibliography}

\begin{thebibliography}{10}
\expandafter\ifx\csname url\endcsname\relax
  \def\url#1{\texttt{#1}}\fi
\expandafter\ifx\csname urlprefix\endcsname\relax\def\urlprefix{URL }\fi
\expandafter\ifx\csname href\endcsname\relax
  \def\href#1#2{#2} \def\path#1{#1}\fi

\bibitem{nielsen_innovation_2022}
J.~Nielsen, C.~B. Tillegreen, D.~Petranovic, Innovation trends in industrial biotechnology, Trends in Biotechnology 40~(10) (2022) 1160--1172.
\newblock \href {https://doi.org/10.1016/j.tibtech.2022.03.007} {\path{doi:10.1016/j.tibtech.2022.03.007}}.

\bibitem{ko_tools_2020}
Y.-S. Ko, J.~W. Kim, J.~A. Lee, T.~Han, G.~B. Kim, J.~E. Park, S.~Y. Lee, Tools and strategies of systems metabolic engineering for the development of microbial cell factories for chemical production, Chemical Society Reviews 49~(14) (2020) 4615--4636.
\newblock \href {https://doi.org/10.1039/D0CS00155D} {\path{doi:10.1039/D0CS00155D}}.

\bibitem{hartline_dynamic_2021}
C.~J. Hartline, A.~C. Schmitz, Y.~Han, F.~Zhang, Dynamic control in metabolic engineering: theories, tools, and applications, Metabolic Engineering 63 (2021) 126--140.
\newblock \href {https://doi.org/10.1016/j.ymben.2020.08.015} {\path{doi:10.1016/j.ymben.2020.08.015}}.

\bibitem{tian_refactoring_2020}
R.~Tian, G.~Du, Y.~Liu, Refactoring and optimization of metabolic network, in: Systems and {Synthetic} {Metabolic} {Engineering}, Elsevier, 2020, pp. 77--105.
\newblock \href {https://doi.org/10.1016/B978-0-12-821753-5.00004-6} {\path{doi:10.1016/B978-0-12-821753-5.00004-6}}.

\bibitem{MAO2024108401}
J.~Mao, H.~Zhang, Y.~Chen, L.~Wei, J.~Liu, J.~Nielsen, Y.~Chen, N.~Xu, Relieving metabolic burden to improve robustness and bioproduction by industrial microorganisms, Biotechnology Advances 74 (2024) 108401.
\newblock \href {https://doi.org/10.1016/j.biotechadv.2024.108401} {\path{doi:10.1016/j.biotechadv.2024.108401}}.

\bibitem{JIANG20231430}
Y.~Jiang, R.~Wu, W.~Zhang, F.~Xin, M.~Jiang, Construction of stable microbial consortia for effective biochemical synthesis, Trends in Biotechnology 41~(11) (2023) 1430--1441.
\newblock \href {https://doi.org/10.1016/j.tibtech.2023.05.008} {\path{doi:10.1016/j.tibtech.2023.05.008}}.

\bibitem{darvishi_applications_2024}
F.~Darvishi, S.~Rafatiyan, M.~H. Abbaspour Motlagh~Moghaddam, E.~Atkinson, R.~Ledesma-Amaro, Applications of synthetic yeast consortia for the production of native and non-native chemicals, Critical Reviews in Biotechnology 44~(1) (2024) 15--30.
\newblock \href {https://doi.org/10.1080/07388551.2022.2118569} {\path{doi:10.1080/07388551.2022.2118569}}.

\bibitem{astrom_feedback_2021}
K.~J. Åström, R.~M. Murray, Feedback systems: an introduction for scientists and engineers, 2nd Edition, Princeton University Press, Princeton, 2021.

\bibitem{JONES2023209}
J.~Jones, D.~Kindembe, H.~Branton, N.~Lawal, E.~L. Montero, J.~Mack, S.~Shi, R.~Patton, G.~Montague, Improved control strategies for the environment within cell culture bioreactors, Food and Bioproducts Processing 138 (2023) 209--220.
\newblock \href {https://doi.org/10.1016/j.fbp.2023.02.004} {\path{doi:10.1016/j.fbp.2023.02.004}}.

\bibitem{zupke_mpc}
C.~Zupke, L.~J. Brady, P.~G. Slade, P.~Clark, R.~G. Caspary, B.~Livingston, L.~Taylor, K.~Bigham, A.~E. Morris, R.~W. Bailey, Real-time product attribute control to manufacture antibodies with defined n-linked glycan levels, Biotechnology Progress 31~(5) (2015) 1433--1441.
\newblock \href {https://doi.org/10.1002/btpr.2136} {\path{doi:10.1002/btpr.2136}}.

\bibitem{CRAVEN2014344}
S.~Craven, J.~Whelan, B.~Glennon, Glucose concentration control of a fed-batch mammalian cell bioprocess using a nonlinear model predictive controller, Journal of Process Control 24~(4) (2014) 344--357.
\newblock \href {https://doi.org/10.1016/j.jprocont.2014.02.007} {\path{doi:10.1016/j.jprocont.2014.02.007}}.

\bibitem{espinel_met_cyberg}
S.~Espinel-Ríos, B.~Morabito, J.~Pohlodek, K.~Bettenbrock, S.~Klamt, R.~Findeisen, Toward a modeling, optimization, and predictive control framework for fed-batch metabolic cybergenetics, Biotechnology and Bioengineering 121~(1) (2024) 366--379.
\newblock \href {https://doi.org/10.1002/bit.28575} {\path{doi:10.1002/bit.28575}}.

\bibitem{espinel-rios_hybrid_2024}
S.~Espinel-Ríos, J.~L. Avalos, Hybrid physics-informed metabolic cybergenetics: process rates augmented with machine-learning surrogates informed by flux balance analysis, Industrial \& Engineering Chemistry Research 63~(15) (2024) 6685--6700.
\newblock \href {https://doi.org/10.1021/acs.iecr.4c00001} {\path{doi:10.1021/acs.iecr.4c00001}}.

\bibitem{rawlings_model_2020}
J.~B. Rawlings, D.~Q. Mayne, M.~Diehl, Model predictive control: theory, computation, and design, 2nd Edition, Nob Hill Publishing, Santa Barbara, California, 2020.

\bibitem{ADETOLA2009320}
V.~Adetola, D.~DeHaan, M.~Guay, Adaptive model predictive control for constrained nonlinear systems, Systems \& Control Letters 58~(5) (2009) 320--326.
\newblock \href {https://doi.org/10.1016/j.sysconle.2008.12.002} {\path{doi:10.1016/j.sysconle.2008.12.002}}.

\bibitem{JABARIVELISDEH2020106744}
B.~Jabarivelisdeh, L.~Carius, R.~Findeisen, S.~Waldherr, Adaptive predictive control of bioprocesses with constraint-based modeling and estimation, Computers \& Chemical Engineering 135 (2020) 106744.
\newblock \href {https://doi.org/10.1016/j.compchemeng.2020.106744} {\path{doi:10.1016/j.compchemeng.2020.106744}}.

\bibitem{petsagkourakis_reinforcement_2020}
P.~Petsagkourakis, I.~Sandoval, E.~Bradford, D.~Zhang, E.~Del Rio-Chanona, Reinforcement learning for batch bioprocess optimization, Computers \& Chemical Engineering 133 (2020) 106649.
\newblock \href {https://doi.org/10.1016/j.compchemeng.2019.106649} {\path{doi:10.1016/j.compchemeng.2019.106649}}.

\bibitem{espinel-rios_enhancing_2024}
S.~Espinel-Ríos, J.~Q. Mo, D.~Zhang, E.~A. del Rio-Chanona, J.~L. Avalos, Enhancing reinforcement learning for population setpoint tracking in co-cultures, arXiv (2024).
\newblock \href {https://doi.org/10.48550/ARXIV.2411.09177} {\path{doi:10.48550/ARXIV.2411.09177}}.

\bibitem{sutton_reinforcement_2018}
R.~S. Sutton, A.~G. Barto, Reinforcement learning: an introduction, 2nd Edition, Adaptive computation and machine learning series, The MIT Press, Cambridge, Massachusetts, 2018.

\bibitem{dong_introduction_2020}
Z.~Ding, Y.~Huang, H.~Yuan, H.~Dong, Introduction to reinforcement learning, in: H.~Dong, Z.~Ding, S.~Zhang (Eds.), Deep {Reinforcement} {Learning}, Springer Singapore, Singapore, 2020, pp. 47--123.
\newblock \href {https://doi.org/10.1007/978-981-15-4095-0_2} {\path{doi:10.1007/978-981-15-4095-0_2}}.

\bibitem{hilo_mpc}
J.~Pohlodek, B.~Morabito, C.~Schlauch, P.~Zometa, R.~Findeisen, Flexible development and evaluation of machine‐learning‐supported optimal control and estimation methods via {HILO}‐{MPC}, International Journal of Robust and Nonlinear Control (2024) rnc.7275\href {https://doi.org/10.1002/rnc.7275} {\path{doi:10.1002/rnc.7275}}.

\bibitem{zhang_deep_2023}
J.~Zhang, C.~Zhao, J.~Ding, Deep reinforcement learning with domain randomization for overhead crane control with payload mass variations, Control Engineering Practice 141 (2023) 105689.
\newblock \href {https://doi.org/10.1016/j.conengprac.2023.105689} {\path{doi:10.1016/j.conengprac.2023.105689}}.

\bibitem{NIPS1999_464d828b}
R.~S. Sutton, D.~McAllester, S.~Singh, Y.~Mansour, Policy gradient methods for reinforcement learning with function approximation, in: S.~Solla, T.~Leen, K.~M\"{u}ller (Eds.), Advances in Neural Information Processing Systems, Vol.~12, MIT Press, 1999.

\bibitem{liu_how_2017}
S.~Liu, How cells grow, in: Bioprocess {Engineering}, Elsevier, 2017, pp. 629--697.
\newblock \href {https://doi.org/10.1016/B978-0-444-63783-3.00011-3} {\path{doi:10.1016/B978-0-444-63783-3.00011-3}}.

\bibitem{jayaraman_blue_2016}
P.~Jayaraman, K.~Devarajan, T.~K. Chua, H.~Zhang, E.~Gunawan, C.~L. Poh, Blue light-mediated transcriptional activation and repression of gene expression in bacteria, Nucleic Acids Research 44~(14) (2016) 6994--7005.
\newblock \href {https://doi.org/10.1093/nar/gkw548} {\path{doi:10.1093/nar/gkw548}}.

\bibitem{multamaki_optogenetic_2022}
E.~Multamäki, A.~García~de Fuentes, O.~Sieryi, A.~Bykov, U.~Gerken, A.~T. Ranzani, J.~Köhler, I.~Meglinski, A.~Möglich, H.~Takala, Optogenetic control of bacterial expression by red light, ACS Synthetic Biology 11~(10) (2022) 3354--3367.
\newblock \href {https://doi.org/10.1021/acssynbio.2c00259} {\path{doi:10.1021/acssynbio.2c00259}}.

\bibitem{SENN1994424}
H.~Senn, U.~Lendenmann, M.~Snozzi, G.~Hamer, T.~Egli, The growth of {Escherichia coli} in glucose-limited chemostat cultures: a re-examination of the kinetics, Biochimica et Biophysica Acta (BBA) - General Subjects 1201~(3) (1994) 424--436.
\newblock \href {https://doi.org/10.1016/0304-4165(94)90072-8} {\path{doi:10.1016/0304-4165(94)90072-8}}.

\bibitem{hadicke_ecolicore2_2017}
O.~Hädicke, S.~Klamt, {EColiCore2}: a reference network model of the central metabolism of {Escherichia} coli and relationships to its genome-scale parent model, Scientific Reports 7~(1) (2017) 39647.
\newblock \href {https://doi.org/10.1038/srep39647} {\path{doi:10.1038/srep39647}}.

\bibitem{milo_cell_2016}
R.~Milo, R.~Phillips, Cell biology by the numbers, Garland Science, Taylor \& Francis Group, New York, NY, 2016.

\bibitem{pytorch}
A.~Paszke, S.~Gross, F.~Massa, A.~Lerer, J.~Bradbury, G.~Chanan, T.~Killeen, Z.~Lin, N.~Gimelshein, L.~Antiga, A.~Desmaison, A.~K\"{o}pf, E.~Yang, Z.~DeVito, M.~Raison, A.~Tejani, S.~Chilamkurthy, B.~Steiner, L.~Fang, J.~Bai, S.~Chintala, PyTorch: an imperative style, high-performance deep learning library, Curran Associates Inc., Red Hook, NY, USA, 2019.

\bibitem{Andersson2019}
J.~A.~E. Andersson, J.~Gillis, G.~Horn, J.~B. Rawlings, M.~Diehl, {CasADi}: a software framework for nonlinear optimization and optimal control, Mathematical Programming Computation 11~(1) (2019) 1--36.
\newblock \href {https://doi.org/10.1007/s12532-018-0139-4} {\path{doi:10.1007/s12532-018-0139-4}}.

\end{thebibliography}

\end{document}